%% file: main.tex
\documentclass[fleqn,usenatbib]{mnras}

\usepackage{newtxtext,newtxmath}

\usepackage[T1]{fontenc}

\DeclareRobustCommand{\VAN}[3]{#2}
\let\VANthebibliography\thebibliography
\def\thebibliography{\DeclareRobustCommand{\VAN}[3]{##3}\VANthebibliography}

\usepackage{graphicx}	
\usepackage{amsmath}	

\usepackage{subfig}

\DeclareCaptionLabelFormat{continued}{Figure #1~#2 (Continued)}
\title{ML-MOC: Machine Learning (kNN and GMM) based Membership Determination for Open Clusters}

\author[Manan Agarwal et al.]{
Manan Agarwal,$^{1}$\thanks{E-mail: f2016607@pilani.bits-pilani.ac.in}
Khushboo K. Rao,$^{1}$
Kaushar Vaidya,$^{1}$
and Souradeep Bhattacharya$^{2}$
\\
$^{1}$Department of Physics, Birla Institute of Technology and Science – Pilani, 333031 Rajasthan, India\\
$^{2}$Inter University Centre for Astronomy and Astrophysics, Ganeshkhind, Post Bag 4, Pune 411007, India
}

\date{Accepted XXX. Received YYY; in original form ZZZ}

\pubyear{2021}

\begin{document}
\label{firstpage}
\pagerange{\pageref{firstpage}--\pageref{lastpage}}
\maketitle

\begin{abstract}
The existing open cluster membership determination algorithms are either prior dependent on some known parameters of clusters or are not automatable to large samples of clusters. In this paper, we present, ML-MOC, a new machine learning based approach to identify likely members of open clusters using the Gaia DR2 data, and no a priori information about cluster parameters. We use the k-Nearest Neighbours (kNN) algorithm and the Gaussian Mixture Model (GMM) on the high-precision proper motions and parallax measurements from Gaia DR2 data to determine the membership probabilities of individual sources down to $G\sim$20 mag. To validate the developed method, we apply it on fifteen open clusters: M67, NGC 2099, NGC 2141, NGC 2243, NGC 2539, NGC 6253, NGC 6405, NGC 6791, NGC 7044, NGC 7142, NGC 752, Blanco 1, Berkeley 18, IC 4651, and Hyades. These clusters differ in terms of their ages, distances, metallicities, extinctions and cover a wide parameter space in proper motions and parallaxes with respect to the field population. The extracted members produce clean colour-magnitude diagrams and our astrometric parameters of the clusters are in good agreement with the values derived by the previous works. The estimated degree of contamination in the extracted members range between 2$\%$ and 12$\%$. The results show that ML-MOC is a reliable approach to segregate the open cluster members from the field stars.
\end{abstract}




\input{chapters/introduction}

\input{chapters/data}

\input{chapters/methodology}

\input{chapters/results}

\input{chapters/conclusions}

\section*{Acknowledgements}

The authors are grateful to the anonymous referee for the valuable comments. This work has made use of data from the European Space Agency (ESA) mission Gaia (\url{https: //www.cosmos.esa.int/gaia}), processed by the Gaia Data Processing and Analysis Consortium (DPAC, \url{https://www.cosmos.esa.int/web /gaia/dpac/consortium}). Funding for the DPAC has been provided by national institutions, in particular the institutions participating in the Gaia Multilateral Agreement. This research has made use of the VizieR catalogueue access tool, CDS, Strasbourg, France. This research made use of Astropy, a community-developed core Python package for Astronomy \citep{astropy:2013, astropy:2018}, scikit-learn \citep{scikit-learn} and Numpy \citep{numpy}. The figures in this paper were produced with Bokeh, a Python library for interactive visualization \citep{Bokeh} and Matplotlib \citep{Matplotlib}. This research also made use of NASA’s Astrophysics Data System (ADS).

\section*{Data Availability}

The data underlying this article are publicly available at \url{https://archives.esac.esa.int/gaia}. The derived data generated in this research will be shared on reasonable request to the corresponding author.




\bibliographystyle{mnras}
\bibliography{references} 



\appendix

\input{appendices/Be18}
\input{appendices/Hyades}
\input{appendices/additional_figures}

\bsp	
\label{lastpage}
\end{document}

%% file: chapters/introduction.tex
\section{Introduction}
\label{sec:introduction}

Galactic or Open Clusters (OCs) are the ideal laboratories to study the formation and evolution of stars \citep{Krumholz2019}, as they provide us chemically homogeneous groups of stars that are of the same age, share the same kinematics (proper motion and radial velocity), and are located at approximately the same given distance from us. The vast majority of them are located close to the Galactic plane and thus serve as excellent tracers of the recent formation history of the Galactic disk \citep[e.g.][]{Friel1995,Chen2003,Jacobson2016}. Accurate determination of cluster membership is essential for the studies of open clusters as it directly influences the estimation of the fundamental astrophysical parameters of clusters, e.g. age, photometric distance, reddening, and metallicity, among other things. The well-known open cluster catalogues, \cite{Dias2002} and \citet[][K13 hereafter]{Kharchenko2013}, list about 3000 open clusters.

The ongoing European Space Agency (ESA) mission Gaia \citep{Perryman2001}, in particular the second Gaia Data Release, Gaia DR2 \citep{GaiaCollab}, has revolutionized the studies of open clusters by providing astrometric measurements with unprecedented precision. \citet[][CG18 hereafter]{Cantat2018} used Gaia DR2 to compute the membership probabilities for 1229 OCs including 60 previously unknown OCs. Gaia DR2 also facilitated the studies aimed at finding new OCs. \cite{Sim2019} reported 207 OCs by visually inspecting proper motion diagrams; \cite{Liu2019} further found 76 OCs that were previously unknown; \cite{Castro2019} identified 53 new OCs near the Galactic anticentre. \cite{Cantat-Gaudin2020} included the membership probabilities for the newly discovered OCs to extend their catalogue to a total of 1481 OCs. \cite{Castro2020} recently reported 582 new OCs in the Galactic disk. As illustrated by \cite{Hunt}, the search for new OCs using the Gaia data is far from over.

Various methods have been used for membership determination based on the analysis of the positions, proper motions, parallaxes, radial velocities, photometry, and their combinations \citep{Vasilevskis1958,Sanders1971,Cabrera-Cano1990,Zhao1990,Galadi1998,Clusterix}. In the last few years, machine learning algorithms such as, DBSCAN \citep{GaoDBSCAN,SouraDBSCAN}, HDBSCAN \citep{Hunt}, KMeans \citep{Aziz2016}, GMM \citep{GaoGMM}, Random Forest \citep{GaoNGC6405}, UPMASK \citep{Cantat2018}, and Artificial Neural Network \citep{Castro-Ginard2018}, have been put to the task of separating the true members of open clusters from the field stars. Most of these previous studies have only been applied to a few old open clusters (e.g. \citealt{GaoDBSCAN} studied NGC 188; \citealt{Aziz2016} studied NGC 188 and NGC 2266) or are highly sensitive to the initial sample selection, making them uneasy for being scalable (e.g. \citealt{GaoNGC6405} and \citealt{GaoGMM}). Methods developed by \cite{Cantat2018} and \cite{Castro-Ginard2018} were applied on a large number of open clusters, but they limited their membership analysis to sources brighter than $G \sim$18 mag and $G \sim$17 mag, respectively. Furthermore, \cite{Cantat2018} needs a priori information (distance and radius) about the cluster. \cite{Castro-Ginard2018}, on the other hand, does not obtain the membership probability for individual stars.

In this paper, we propose, ML-MOC, a new probabilistic membership determination algorithm for open cluster members down to $G\sim$20 mag using only Gaia DR2. Our algorithm is based on k-Nearest Neighbours algorithm (kNN, \citealt{kNN}) and Gaussian Mixture Model (GMM, \citealt{GMM}). We apply it on the high-precision Gaia DR2 proper motions and parallaxes to determine the membership probability of individual sources. We aim to facilitate homogeneous analysis of open cluster populations by developing a robust algorithm that works reliably on a large number of open clusters. The method is applied to fifteen open clusters: M67, NGC 2099, NGC 2141, NGC 2243, NGC 2539, NGC 6253, NGC 6405, NGC 6791, NGC 7044, NGC 7142, NGC 752, Blanco 1, Berkeley 18, IC 4651, and Hyades, which cover a range of ages, distances, metallicities and extinction.

The remainder of this paper is organized as follows. Section~\ref{sec:data} describes the Gaia DR2 data and the cluster sample studied in this work, Section~\ref{sec:methodology} describes the methodology of sample selection and the membership determination algorithm using M67 as the example cluster, in Section~\ref{sec:results}, we present our estimates of the degree of contamination, and make a comparison between our extracted members and the previously identified members of the clusters in the literature. Section~\ref{sec:conclusions} summarizes the main conclusions of this work.

%% file: chapters/data.tex
\section{Data and Cluster Sample}
\label{sec:data}

We use Gaia DR2 \citep{GaiaCollab} which catalogues more than 1.3 billion sources with unprecedented astrometric precision and accuracy. It provides a five parameter astrometric solution ($\alpha, \delta, \mu_{\alpha*}, \mu_{\delta}, \omega$): stellar positions RA ($\alpha$) and DEC ($\delta$), proper motions in RA ($\mu_{\alpha*}$) and in DEC ($\mu_{\delta}$), and parallaxes ($\omega$). It provides photometry for three broad bands ($G$, $G_{\mathrm{BP}}$, and $G_{\mathrm{RP}}$), containing sources up to a limiting magnitude of $G\sim$21 mag. The large magnitude range leads to significant differences in the precision of various parameters of the bright and the faint sources. In parallax measurements, the uncertainties reach a precision of 0.02 milliarcsecond (mas hereafter) for $G < 14$ mag sources, and 2 mas for sources near $G\sim$21 mag. In proper motions measurements, the uncertainties range between 0.05 mas yr$^{-1}$ for $G < 14$ mag sources, and 5 mas yr$^{-1}$ for sources with $G\sim$21 mag. Among the other data products, Gaia DR2 also contains radial velocities for most sources brighter than $G\sim$13 mag.

In this work, we determine the membership probabilities of fifteen open clusters: M67, NGC 2099, NGC 2141, NGC 2243, NGC 2539, NGC 6253, NGC 6405, NGC 6791, NGC 7044, NGC 7142, NGC 752, Blanco 1, Berkeley 18, IC 4651, and Hyades, that are located at different latitudes and cover a wide parameter space in proper motions and parallaxes relative to the foreground and the background contamination. They vary in terms of their ages, 0.53 Gyr to 8.89 Gyr, distances, $\sim$47 pc to $\sim$5500 pc, metallicities [Fe/H], $-$0.54 to 0.43, and suffer from little extinction to as much as $A_V =$ 1.7 mag. We use the cluster M67 (NGC 2682) to demonstrate our methodology, as it is a well studied open cluster and its members are publicly available \citep{M67_1977,M67_2004,M67_2009,M67_WOCS}.

%% file: chapters/methodology.tex
\section{Methodology}
\label{sec:methodology}

 Our approach to membership determination only uses astrometric measurements from Gaia DR2 and does not require any a priori information about the cluster parameters. The method is independent of the cluster density profile and its structure in the photometric space, but relies on the assumption that the field and the cluster can be modelled as two distinct Gaussians in the ($\mu_{\alpha*}$, $\mu_\delta$, $\omega$) space. The extraction of the cluster members is done in three main stages: extracting the sample sources, identifying high probability member sources and extending the list of members by identifying likely moderate probability members.

In the first stage, we use the kNN algorithm and extract the appropriate sample sources by removing the obvious field noise. In the second stage, we normalise proper motions and parallaxes and apply a three-dimensional GMM on the sample sources to identify member stars. Lastly, in the third stage, we include member stars with moderate probability to our selection of cluster members. Following is a detailed description of each stage.

\subsection{First Stage: Extract the Sample sources}

\begin{figure}
	\includegraphics[width=\columnwidth]{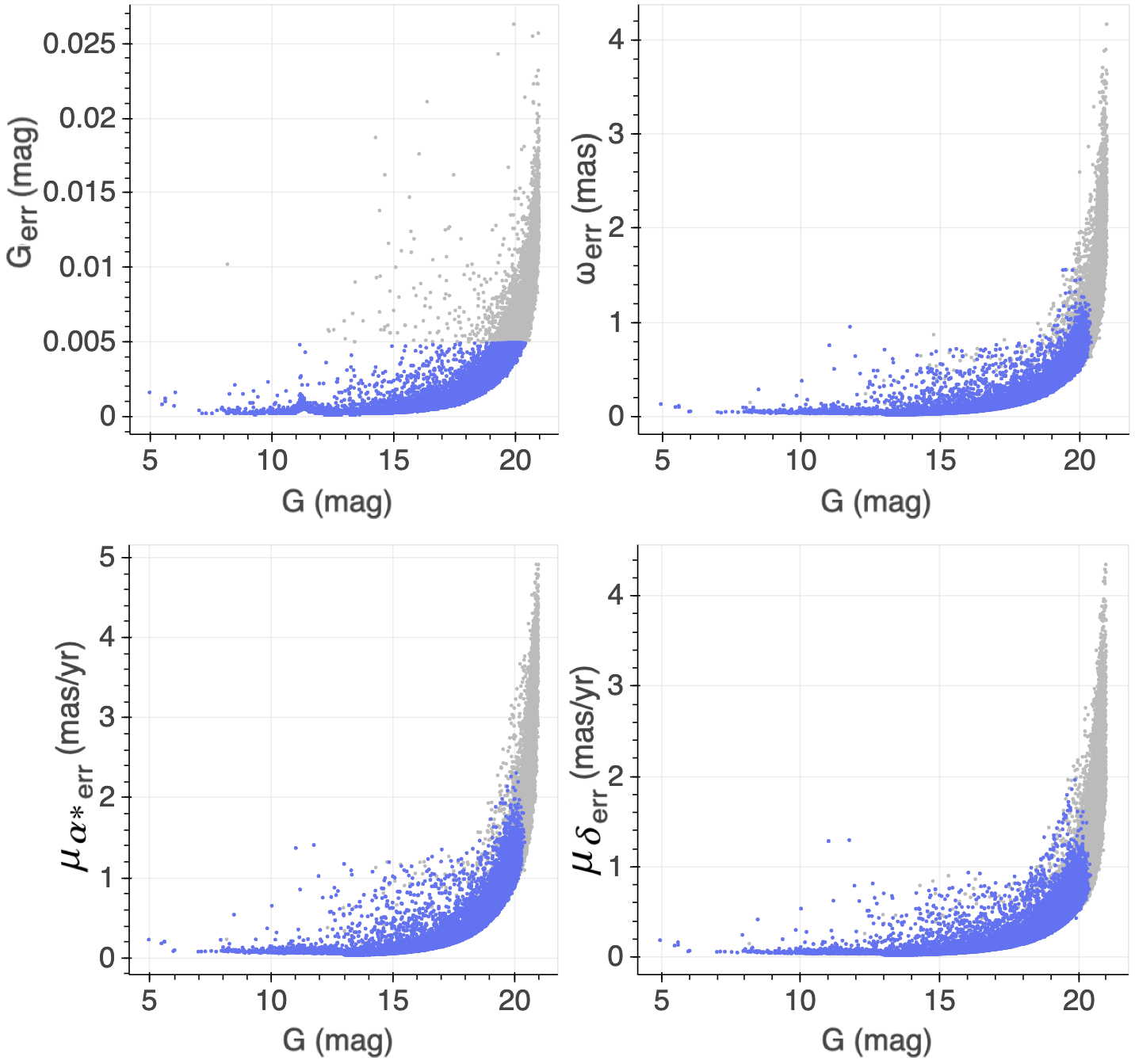}
    \caption{The correlation of errors (photometry, parallax, and proper motions) with the $G$-mag of sources.  The grey points are all Gaia DR2 sources for M67 within a radius of 150 arcmin from the cluster centre.  The sources in blue are those with $G_\mathrm{err}$ < 0.005 mag.}
    \label{fig:M67_error_cut}
\end{figure}

The aim of this stage is to remove a large number of obvious field stars. First, we download sources from Gaia DR2 in a cone around the cluster centre for a value of radius that is greater than the tidal radius of the cluster, as reported in K13. Though our algorithm is quite robust to the choice of this initial radius, as a rule of thumb, we generally use the value of radius that is 1.5 times the tidal radius. For M67, we download sources within a radius of 150 arcmin from the cluster centre. Next, we select the sources which satisfy the following criteria:
\begin{enumerate}
\item{each source must have the five astrometric parameters, positions, proper motions and parallax as well as valid measurements in the three photometric passbands $G, G_{\mathrm{BP}},$ and $G_{\mathrm{RP}}$ in the Gaia DR2 catalog}, 
\item{their parallax values must be non-negative},
\item{the errors in their $G$-mag must be less than 0.005. This last criteria allows us to eliminate sources with high uncertainty while still retaining a fraction of sources down to $G\sim$21 mag.}
\end{enumerate}
The uncertainty in parallaxes depend on the signal-to-noise ratio, which is correlated with the $G$ magnitudes. In Figure~\ref{fig:M67_error_cut}, we show how the criteria (iii) limits the uncertainties of the parallaxes and the proper motions. Based on these three criteria, we get 56135 sources for M67, referred as "\textit{All sources}" hereafter.

\begin{figure}
	\includegraphics[width=\columnwidth]{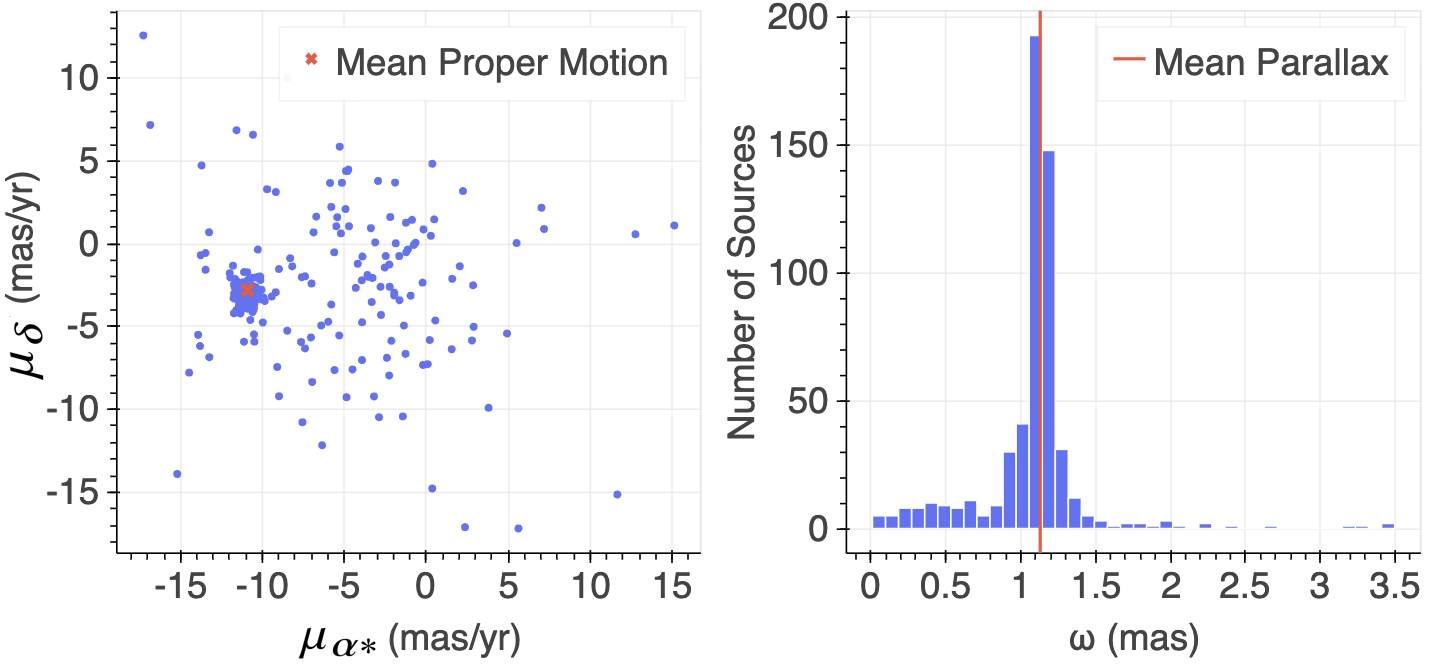}
    \caption{The distributions of proper motions and parallaxes of sources brighter than $G=18$ mag and are lying within 10 arcmin from the cluster centre. The mean of the proper motions and parallaxes, determined by the kNN algorithm, are indicated with a red dot and a red line, respectively.}
    \label{fig:M67_estimates}
\end{figure}

To determine the ranges of proper motions and parallaxes that enclose all the likely cluster members, we estimate the mean proper motions and the mean parallax of the cluster using the kNN \citep{kNN}. It calculates the distance (we use the Euclidean distance) of each data point to its $k$ nearest data points in the feature space \footnote{Feature space refers to the vector space defined by the collection of numerical features used to characterize the data. A data point with n features is represented in an n-dimensional feature space.} for the purpose of classification and regression. It has a time complexity of $O(n$ $log$ $n)$ \citep{kNN_time} and makes no assumption about the underlying data. We used the data point with the least total distance, i.e. the sum of distances to $k$ nearest data points, to make a coarse estimation of the cluster parameters. We applied the kNN algorithm on the 2D proper motion space of these sources to estimate the mean proper motion of the cluster. It should be noted that these are only initial estimates made and used in order to filter out the field stars. Similarly, we performed kNN in the parallax space to estimate the mean parallax ($\overline{\omega}$) and the corresponding distance is obtained from \cite{BailerJones}. Although it is possible to perform kNN on \textit{All sources}, the process will be computationally expensive, time-inefficient, and would have a slim chance of finding the accurate mean parameters for a nearby cluster in the searched region. We thus considered only sources brighter than $G=18$ mag within 10 arcmin from the cluster centre, where we expect the cluster members to dominate the field contamination. For instance, in the case of M67 the kNN step takes only 11.1 milliseconds with the constraints and 602 milliseconds when performed on \textit{All sources}, while not affecting the final membership results. Figure~\ref{fig:M67_estimates} shows the kNN estimates of mean parameters for M67. In Table~\ref{tab:estimates}, we show that these initial estimates of cluster parameters computed by kNN are in good agreement with the corresponding values by CG18.

\input{tables/estimates}

\begin{figure}
	\includegraphics[width=\columnwidth]{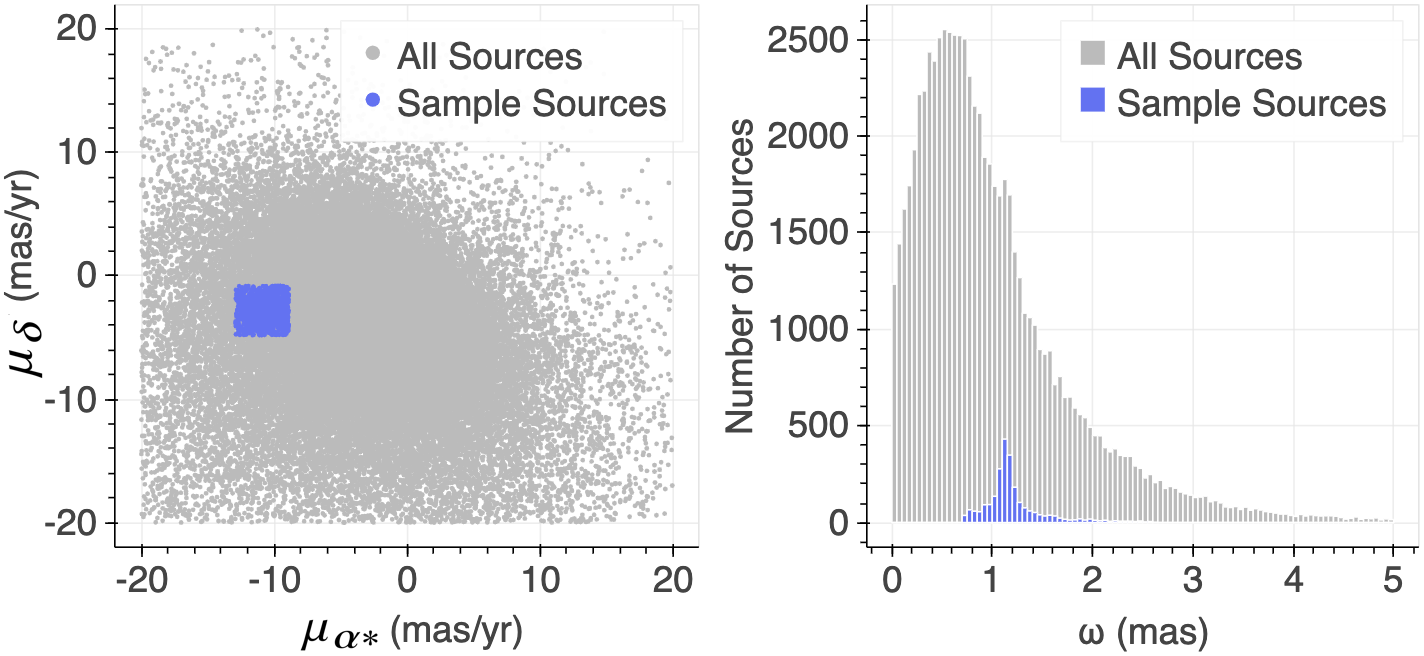}
    \caption{The result of the first-stage analysis. The extracted \textit{Sample sources} for the cluster M67 are shown in blue colour.}
    \label{fig:M67_stage1}
\end{figure}

\begin{figure*} 
	\includegraphics[width=\textwidth]{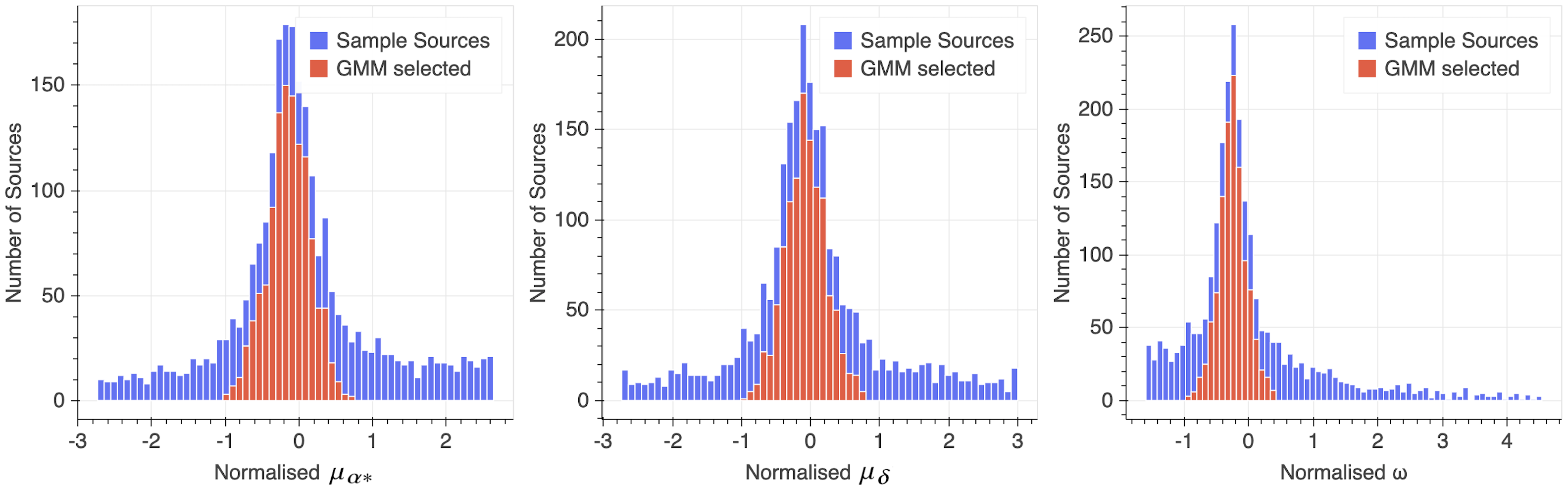}
    \caption{The frequency distributions of normalized proper motions and parallaxes of the \textit{Sample sources} (blue).  The sources with membership probability greater than 0.6, as determined by the GMM, are shown in red bars.}
    \label{fig:M67_stage2}
\end{figure*}

To select the range of proper motions ($\mu_{\alpha*}, \mu_\delta$), we start with a 5 mas/yr wide window and iteratively slide or shrink it, in order to get the mean of the enclosed values close to the mean determined by the kNN. The chosen range of proper motions has a width between 3 and 5 mas/yr. Similarly, we select the range of parallax, starting with a window that has the initial width dependent on the estimated cluster distance, it is wide for nearby clusters and narrow for distant clusters. The selected range of parallax has a width between 0.4 and 2.5 mas \footnote{For clusters which have an overlap in parameter ($\mu_{\alpha*}, \mu_\delta, \omega$) peak of field sources and cluster members (refer Section~\ref{subsec: stellar parameters}), we manually tweak the range.}. For example, as NGC 6405 has an estimated distance of just $\sim$450 pc, we get a 2.5 mas wide parallax range whereas NGC 2243 is estimated at $\sim$4400 pc and we get the parallax range width of 0.4 mas. The sources that are selected by applying the proper motion and parallax range on \textit{All sources} are referred as "\textit{Sample sources}" hereafter. For M67, we retrieve 2427 \textit{Sample sources}. The proper motions and parallaxes of the \textit{Sample sources} in M67 are shown in Figure~\ref{fig:M67_stage1} along with \textit{All sources} of the cluster.

\subsection{Second Stage: Identify the Member sources}


To separate the likely cluster members from the field stars, we apply the GMM \citep{GMM}, an unsupervised clustering algorithm, on the \textit{Sample sources}. Prior to applying the GMM, we perform one more step, i.e., normalize the data ($\mu_{\alpha*}$, $\mu_\delta$, $\omega$) since the similarity measurements are sensitive to the differences in the magnitudes and the scale of the parameters. We perform the standard normalization for the proper motions and parallaxes following the process detailed below. Given N \textit{Sample sources}, each with m parameters $[x^1, x^2, ..., x^m]$, we define the normalized parameter in the $j^{th}$ dimension $X^j_i$ as,

\begin{equation}
    X^j_i=\frac{x^j_i - \mu^j}{\sigma^j} \text{ $(i = 1, 2, ..., N; j = 1, 2, ..., m)$}
\end{equation}

\noindent where $x^j_i$ is the original parameter, $\mu^j$ being the median of $x^j$ distribution, and $\sigma^j$ its standard deviation.

The GMM is a probabilistic model that assumes that all the data points are drawn from a mixture of a finite number of Gaussian distributions with unknown parameters. It can be thought of as an extension of the K-Means clustering \citep{Kmeans_macqueen,Kmeans_lloyd} to incorporate information about not only the mean $(\mu)$, but also the covariance $(\Sigma)$, that describes the ellipsoidal shape of a distribution. It is a fast method with runtime complexity of $O(n)$. The model is fitted by maximizing the likelihood estimates of the distribution parameters using the expectation maximization (EM) algorithm \citep{EMalgo}. Given $N$ data points $x=\{x_1, x_2, x_3, ..., x_N\}$ in a $M$-dimensional parameter space, the $K$-component GMM is defined as

\begin{equation}
    P(x) = \sum_{i=1}^K w_i G( x \mid \mu_i, \Sigma_i), \text{ such that } \sum_{i=1}^K w_i = 1,
\end{equation}

\noindent where $P(x)$ denotes the probability distribution of data points $x$ and $w_i$ is the mixture weight of the $i$th Gaussian component, $G(x\mid \mu_i, \Sigma_i)$, defined as

\begin{equation}
\label{eq:GMM}
    G(x \mid \mu_i, \Sigma_i) = \frac{exp[- \frac{1}{2} (x-\mu_i)^T \Sigma_i^{-1}(x-\mu_i)]}{(2\pi)^{M/2} \sqrt{|\Sigma_i|}}
\end{equation}

\noindent In eq. (\ref{eq:GMM}), $\mu_i$ and $\Sigma_i$ are the mean vector and the full covariance matrix of the $i$th Gaussian component, respectively. The GMM assigns each data point a soft membership probability to each cluster, i.e. how likely the data point is to be described by each cluster. The GMM has been previously used to determine the membership probabilities of the sources of open clusters (M67 by \citealt{Uribe2006} using proper motion; NGC6791 by \citealt{GaoGMM} using 5D astrometry data i.e. position, proper motions and parallax).

\begin{figure*}  
	\includegraphics[width=\textwidth]{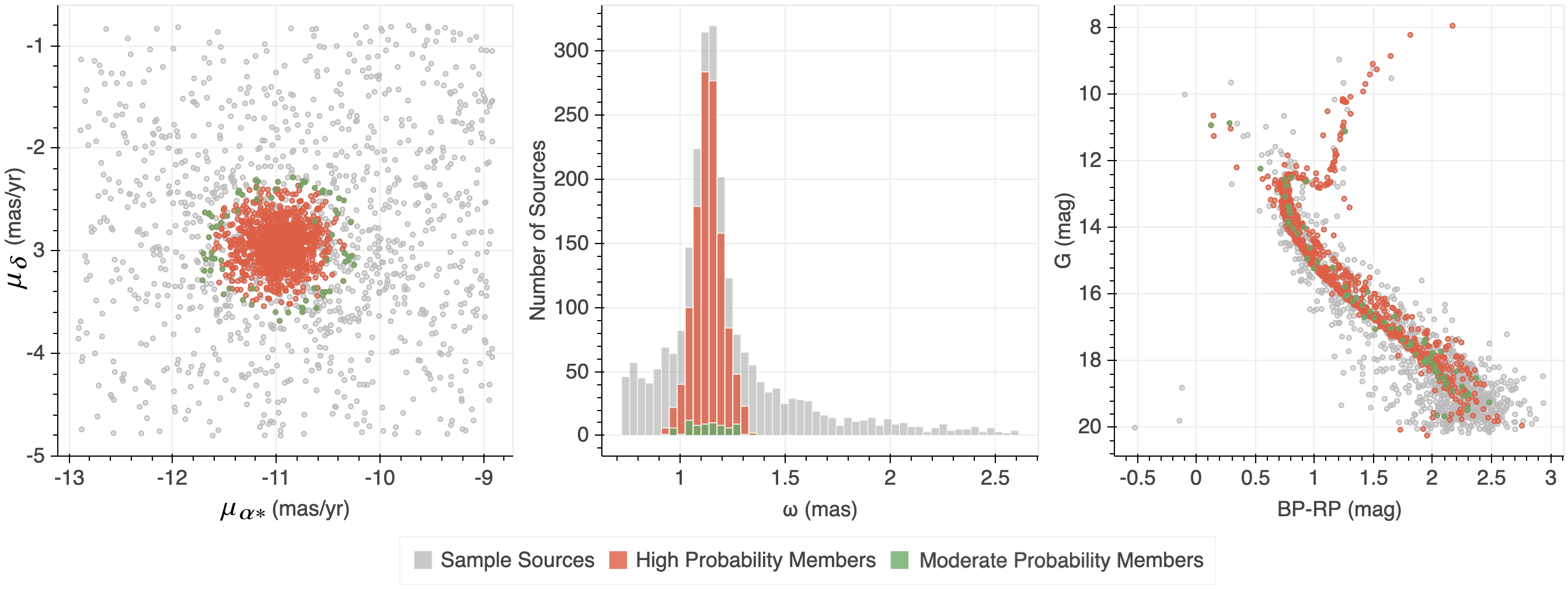}
    \caption{The \textit{Sample sources} of M67 segregated between field stars, high probability members, and moderate probability members.}
    \label{fig:M67_stage3}
\end{figure*}

\begin{figure}  
	\includegraphics[width=\columnwidth]{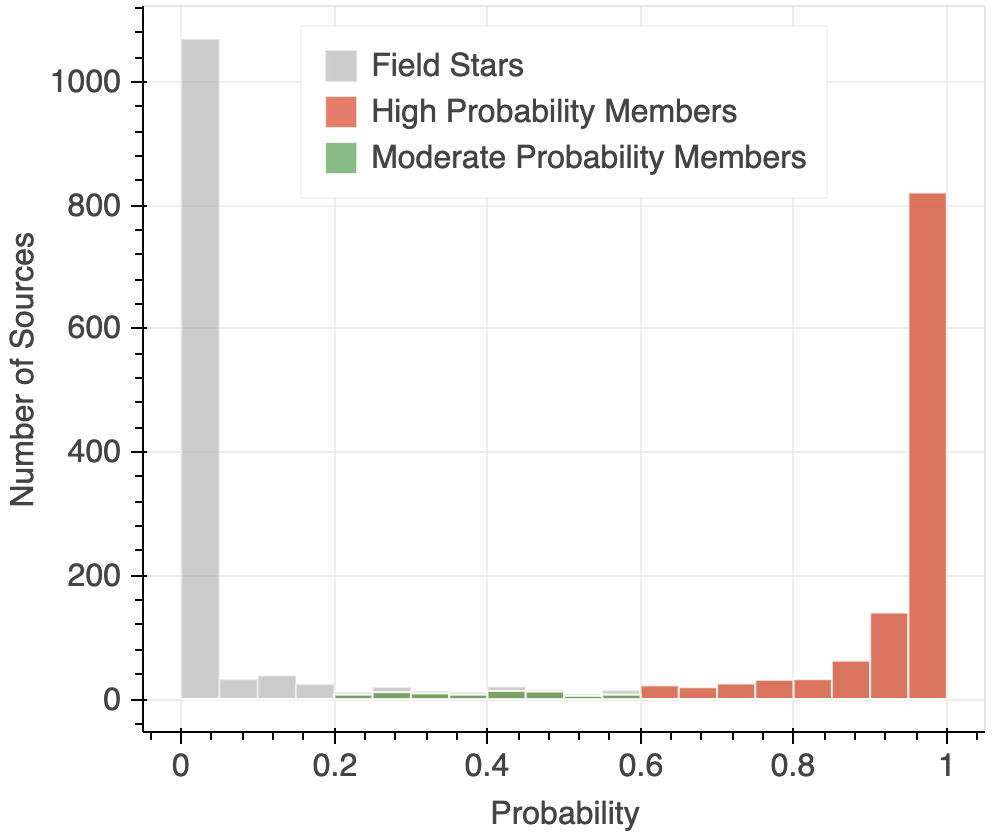}
    \caption{Distribution of the membership probabilities assigned by the GMM to the \textit{Sample sources} of M67.}
    \label{fig:M67_probability}
\end{figure}

In order to get reliable membership probability estimates for cluster members using the GMM, the following conditions should be met: (1) there is a high precision data set, (2) there is a high proportion of cluster members i.e. a high cluster sources to field sources ratio, and (3) there is a difference in the location of the peaks of the distribution of the field star and cluster members \citep{DeGraeve1979}. If even one of these conditions is not met, the computed membership probabilities will lose significance. The unprecedented high-quality astrometry data of Gaia DR2 and the initial selection of sources with the $G$-mag error less than 0.005, ensures that we have a reliable data set to satisfy the first condition. To tackle the second of the above issues, namely the need of high proportion of cluster members in our sample on which we apply the GMM, we choose to apply the GMM on the \textit{Sample sources} instead of \textit{All sources}. The third condition is not in our hand, but it is unlikely that the peak of the field stars coincides with that of the cluster members in all the three dimensions ($\mu_{\alpha*}$, $\mu_\delta$, $\omega$). We give examples of applying ML-MOC to clusters which have overlapping peaks in one or more of the parameters: NGC 2539 has overlapping proper motion in RA ($\mu_{\alpha*}$); IC 4651 has overlap in both proper motion RA and DEC ($\mu_{\alpha*}, \mu_\delta$); NGC 2141 and Berkeley 18 (refer Appendix~\ref{app:Be18}) have overlap in proper motion in RA and parallax ($\mu_{\alpha*}$, $\omega$) with the field sources. The membership determination for these clusters is discussed in detail in Section~\ref{subsec: stellar parameters}.

We use a two-component, i.e. the cluster and field, GMM on the normalised three-dimensional parameter space ($\mu_{\alpha*}$, $\mu_\delta$, $\omega$). This assumes that the proper motions and parallaxes of the \textit{Sample sources} follow a two-component Gaussian distribution. We have not used position in the parameter space as that would have restricted the cluster members within a definite ellipsoidal border, in turn, making it impossible to study the morphology of the cluster and to identify potential escapers. It must also be noted that since the sources were downloaded in a circular region, assuming a uniform field distribution, the peak of the field sources in position space will coincide with the peak of the cluster sources, thus making it difficult for GMM to differentiate between them. Applying GMM on the \textit{Sample sources} of M67 gives us 1150 members with a probability higher than 0.6. In Figure~\ref{fig:M67_stage2}, we show the distribution of these sources in the normalised parameter space.

\subsection{Third Stage: Including Moderate Probability Source}
The high probability members ($\geq0.6$) from the GMM give a definite border to the parallax and proper motion. The moderate membership probability sources lie either on the periphery of the proper motion or on the parallax border. We observed that the sources for which the membership was verified by the radial velocity had a probability greater than 0.8 assigned by the GMM. Thus we used the parallax range determined by our member sources with probability greater than 0.8 to filter the low probability members. We extend the member sources by including the \textit{Sample sources} whose parallax values lie in this range and have a membership probability between 0.2 and 0.6. This allows us to selectively include likely cluster members while keeping the degree of contamination to the minimum. In case of M67, we add 71 sources with a membership probability between 0.2 and 0.6, to get a total of 1221 members. Figure~\ref{fig:M67_stage3} shows the vector point diagram of the proper motions, the parallax distribution and the colour-magnitude diagrams (CMD) for the \textit{Sample sources} of M67. In Figure~\ref{fig:M67_probability}, we show the membership probabilities assigned by GMM to the \textit{Sample sources} of M67. We provide the membership results for the rest of the open clusters in Section~\ref{sec:results}.



%% file: tables/estimates.tex
\begin{table*}
	\caption{Comparison of our initial estimates of mean proper motions, mean parallaxes, and mean distances (using \citealt{BailerJones} determined distances), made by using the kNN algorithm, with the corresponding values for the clusters from CG18. Hyades cluster members were not identified by CG18}.
	\label{tab:estimates}
	\begin{tabular}{l|cccc|cccc}
	\hline
    &\multicolumn{4}{c}{Stage 1 (kNN estimated)} &\multicolumn{4}{c}{CG18} \\
        Clusters &$\mu_{\alpha*}$ &$\mu_\delta$ &$\omega$ &dist &$\mu_{\alpha*}$ &$\mu_\delta$ &$\omega$ &dist \\
        &(mas/yr) &(mas/yr) &(mas) &(pc) &(mas/yr) &(mas/yr) &(mas) &(pc) \\
    \hline
        M67 &-10.913 &-2.801 &1.131 &862.3 &-10.986 &-2.964 &1.135 &859.1 \\
        NGC2099 &1.820 &-5.694 &0.688 &1396.4 &1.924 &-5.648 &0.666 &1438.1 \\
        NGC2141 &-0.045 &-0.725 &0.141 &5296.2 &-0.028 &-0.767 &0.196 &4441.3 \\
        NGC2243 &-1.314 &5.522 &0.226 &3758.3 &-1.279 &5.488 &0.211 &4167.8 \\
        NGC2539 &-2.313 &-0.546 &0.765 &1260.0 &-2.331 &-0.584 &0.754 &1277.4 \\
        NGC6253 &-4.450 &-5.226 &0.579 &1698.5 &-4.537 &-5.280 &0.563 &1689.7 \\
        NGC6405 &-1.216 &-5.784 &2.128 &463.9 &-1.306 &-5.847 &2.172 &454.3 \\
        NGC6791 &-0.417 &-2.239 &0.201 &4279.3 &-0.421 &-2.269 &0.192 &4530.8 \\
        NGC7044 &-5.015 &-5.508 &0.297 &2872.6 &-4.976 &-5.526 &0.273 &3315.6 \\
        NGC7142 &-2.698 &-1.260 &0.394 &2368.0 &-2.747 &-1.288 &0.392 &2376.4 \\
        NGC752 &9.738 &-11.865 &2.240 &441.9 &9.810 &-11.713 &2.239 &441.0 \\
        Blanco1 &18.529 &2.561 &4.183 &237.5 &18.739 &2.602 &4.210 &235.9 \\
        Be18 ** &0.848 &-0.044 &0.203 &3757.0 &0.849 &-0.057 &0.152 &5523.5 \\
        IC4651 &-2.402 &-4.901 &1.082 &901.3 &-2.410 &-5.064 &1.056 &921.3 \\
        Hyades &101.790 &-25.502 &21.350 &46.8 & & & & \\
    \hline
    \end{tabular}

    \footnotesize{** Be18 is discussed in detail in Appendix~\ref{app:Be18}}

\end{table*}

%% file: chapters/results.tex
\section{Analysis and Results}
\label{sec:results}



\subsection{Comparison with spectroscopically confirmed members}

The distributions of field stars and cluster members are overlapped both in parallax and in proper motions, so it is expected that some field stars could have been erroneously extracted as members. In order to analyze the accuracy of our membership determination algorithm, we use the members identified by the WIYN Open Cluster Study \citep[WOCS;][]{WOCS} as a proxy of ground truth. WOCS uses the WIYN 3.5m telescope to make definitive cluster membership measurements via precise (0.5 km/s) radial velocities for proper-motion candidate members. The survey extends down to $G\sim$16.5 mag. Out of the thirteen OCs studied in this work, WOCS has observed M67 \citep{M67_WOCS}, NGC 6791 \citep{NGC6791_WOCS}, NGC 6253 \citep{NGC6253_WOCS} and Hyades \citep{Hyades_WOCS}. For M67, it classified 549 sources as members and 558 sources as non-members. We only consider the WOCS sources for which we found a Gaia counterpart with defined astrometric parameters (positions, proper motions and parallax) and valid measurements in the three photometric passbands ($G, G_{\mathrm{BP}},$ and $G_{\mathrm{RP}}$). We retrieve 486 (88.52$\%$) common members with WOCS, while mislabeling 98 sources as members which are identified as non-members by WOCS. CG18 identified 429 (78.14$\%$) common members while mislabeling 79 sources. \cite{GaoM67}, who also uses Gaia DR2 to identify members for M67, finds 510 (92.89$\%$) common members and misclassifies 101 sources. For NGC 6791, WOCS identifies 111 sources as members and 103 sources as non-members. Both, this work and CG18, retrieve 105 (94.59$\%$) common members with WOCS. We mislabel 9 sources whereas CG18 mislabel 12 sources as members which are classified as non-members by WOCS. For NGC 6253, WOCS identifies 61 sources as members and 22 sources as non-members. We successfully retrieve 48 (78.69$\%$) common members with WOCS and mislabel 3 non-members as members. CG18 identify 44 (72.13$\%$) common members with WOCS and misclassify 1 source. Hyades is discussed in detail in Appendix~\ref{app:Hyades}.

\begin{figure}
	\includegraphics[width=\columnwidth]{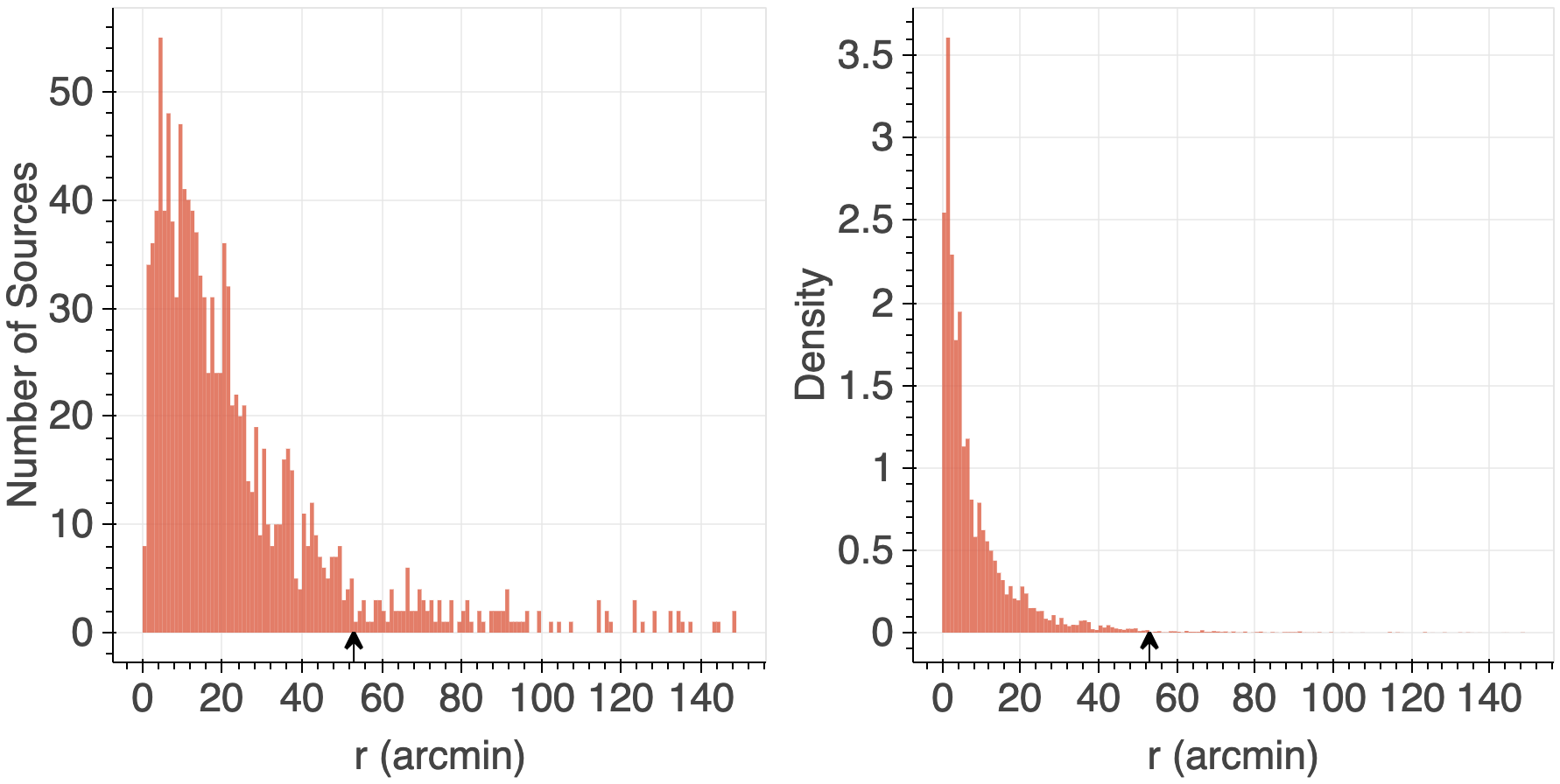}
    \caption{The number and radial density distribution of M67 members. The black arrows mark the cluster radius at 53 arcmin.}
    \label{fig:M67_radial}
\end{figure}

\begin{figure}
	\includegraphics[width=\columnwidth]{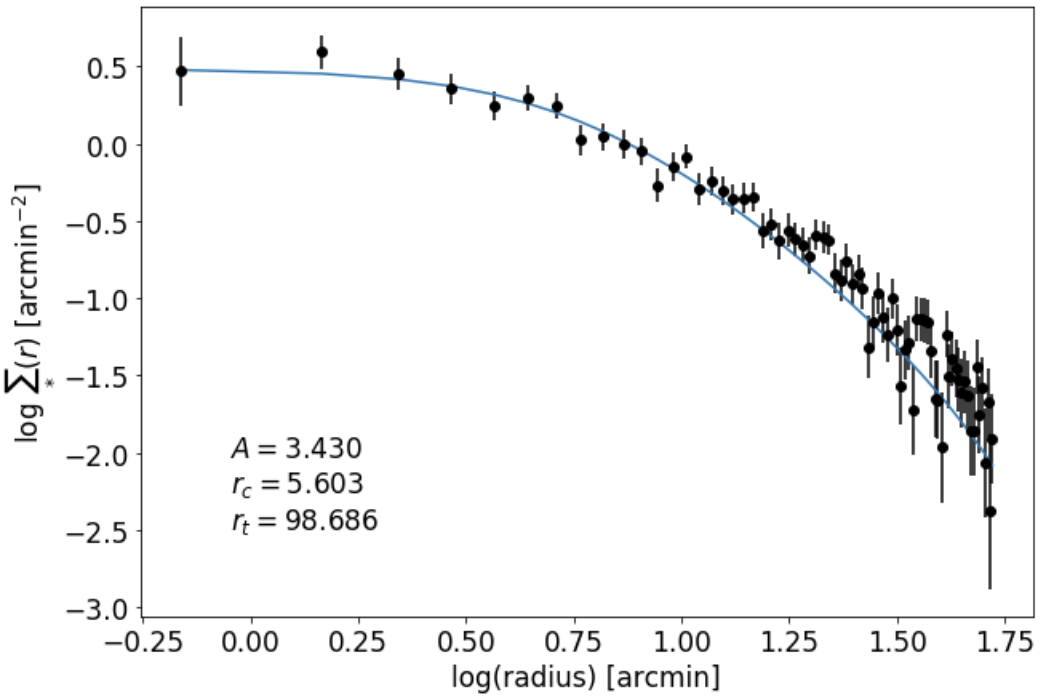}
    \caption{The best fit King's profile of M67 by considering member sources till the cluster radius. The error bars represent $1\sigma$ Poisson errors.}
    \label{fig:M67_kings}
\end{figure}

\input{tables/contamination} 

\begin{figure*}
	\includegraphics[width=\textwidth]{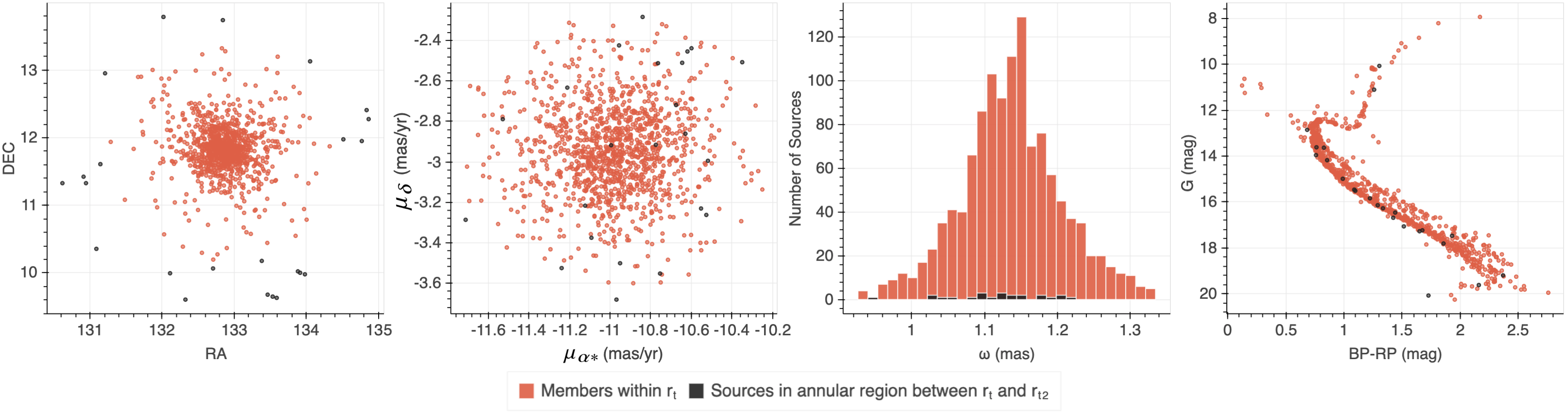}
    \caption{Comparing the member sources (in red) of M67 within $r_t$ with the sources (in black) in the annular region of same area as the area up to $r_t$ from the cluster centre. Out of the 1217 sources within $r_{t2}$ we find 23 sources beyond $r_t$.}
    \label{fig:M67_error}
\end{figure*}

\begin{figure*}
	\includegraphics[width=\textwidth]{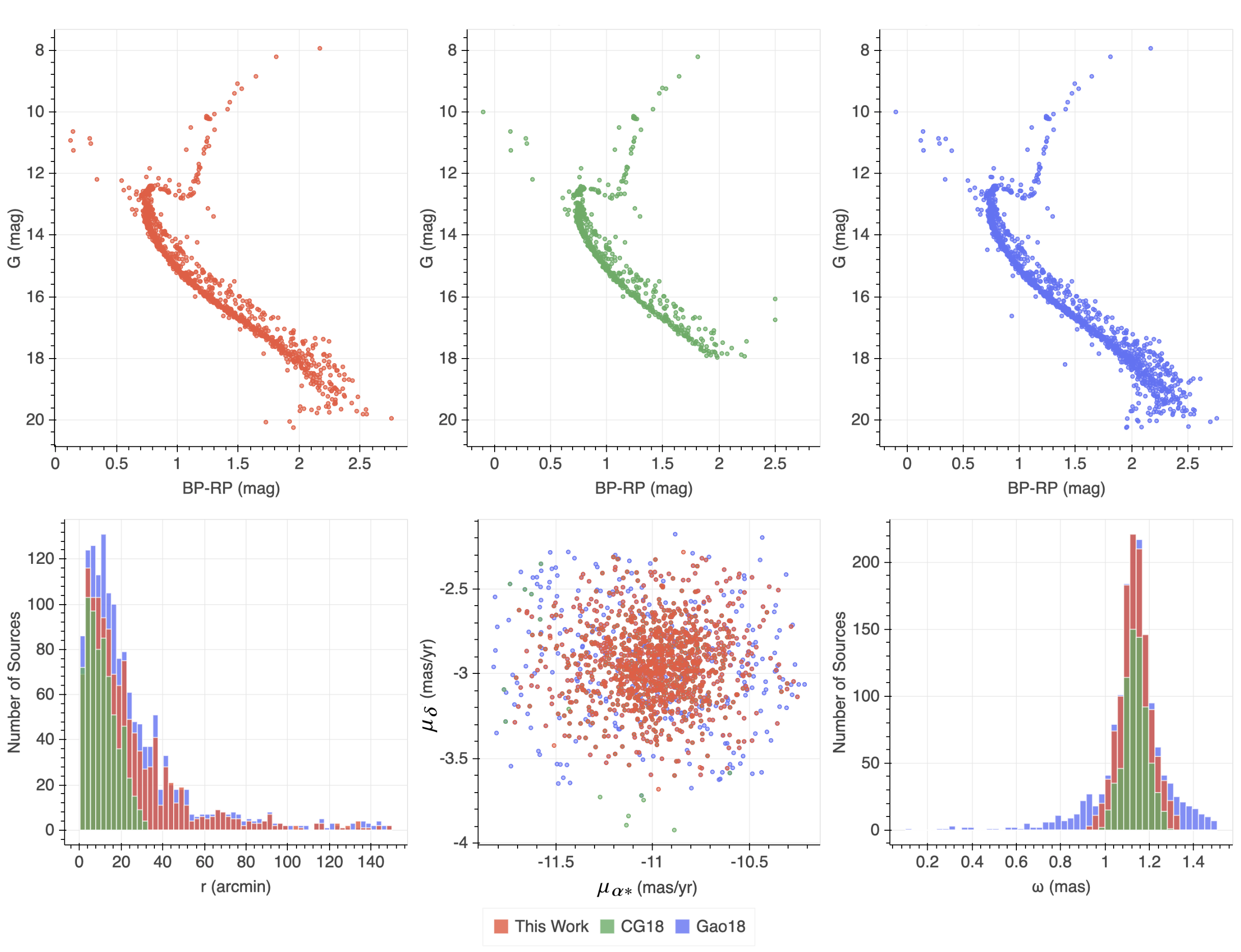}
    \caption{Upper Panel:  The CMDs of members identified for the cluster M67 by the three algorithms.  Lower Panel: The radial distribution, the proper motion distribution, and the parallax distribution of members by the three algorithms.}
    \label{fig:M67_comparison}
\end{figure*}

\subsection{Calculation of cluster radial density parameters}

WOCS data are only available for M67, NGC 6791, NGC 6253 and Hyades, and for sources brighter than $G\sim$16.5 mag. To get a more complete estimate of the degree of contamination, we compare the number of cluster members identified up to the tidal radius ($r_t$) and the radius ($r_{t2}$) that encloses twice the area than the tidal radius. The tidal radius indicates the distance at which the radial density profile reaches the theoretical zero level \citep{King1962}, so all our member sources lying beyond this distance can be classified as noise. To find the tidal radii of the clusters we follow the process described below.

All the identified member sources are used to construct the cluster stellar density radial profiles. We calculated the mean stellar surface density in concentric rings centred on the cluster centre as

\begin{equation}
    \rho_i=\frac{N_i}{\pi(r_{i+1}^2 - r_{i}^2)}
\end{equation}

\noindent where $N_i$ is the number of stars in the $i$-th ring with inner and outer radius $r_i$ and $r_{i+1}$, respectively.

The radial distribution of number of sources, and of projected surface density for M67 are shown in Figure~\ref{fig:M67_radial}. The cluster radius is defined as the distance from the cluster centre, where the combined cluster plus background profile is no longer distinguished from the background alone. We then use all the member sources within the cluster radius to fit the King's profile \citep{King1962} to derive the core ($r_c$) and and tidal radii ($r_t$) of the clusters. Figure~\ref{fig:M67_kings} shows the best fit King's profile for M67.

\begin{figure*}
    \centering
    \subfloat{\includegraphics[width=\textwidth]{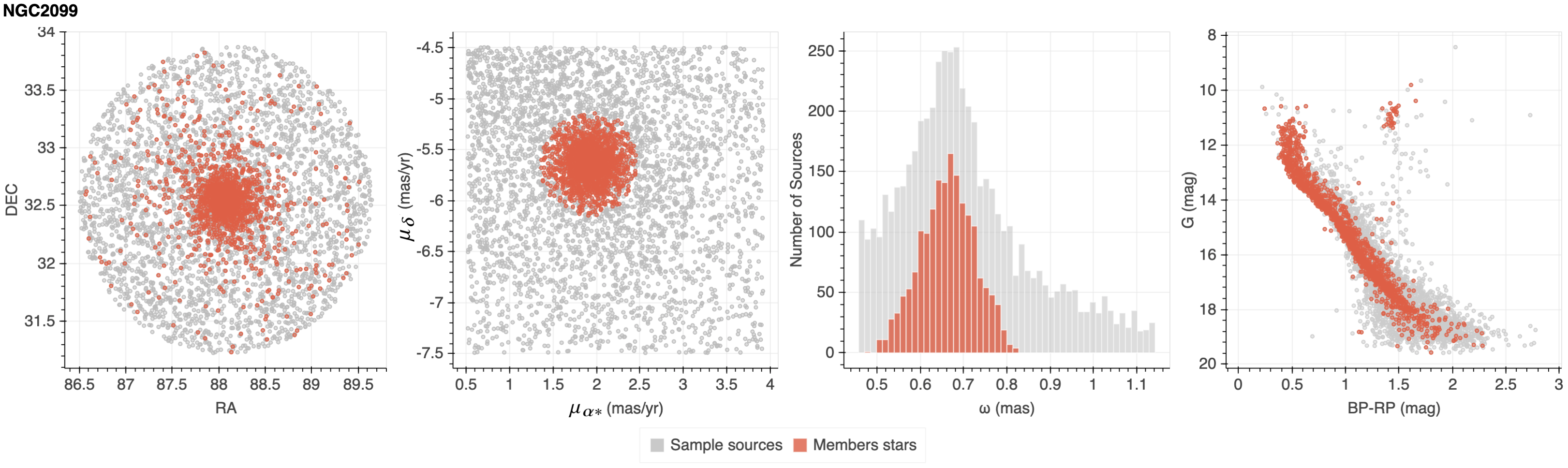}}\\
    \subfloat{\includegraphics[width=\textwidth]{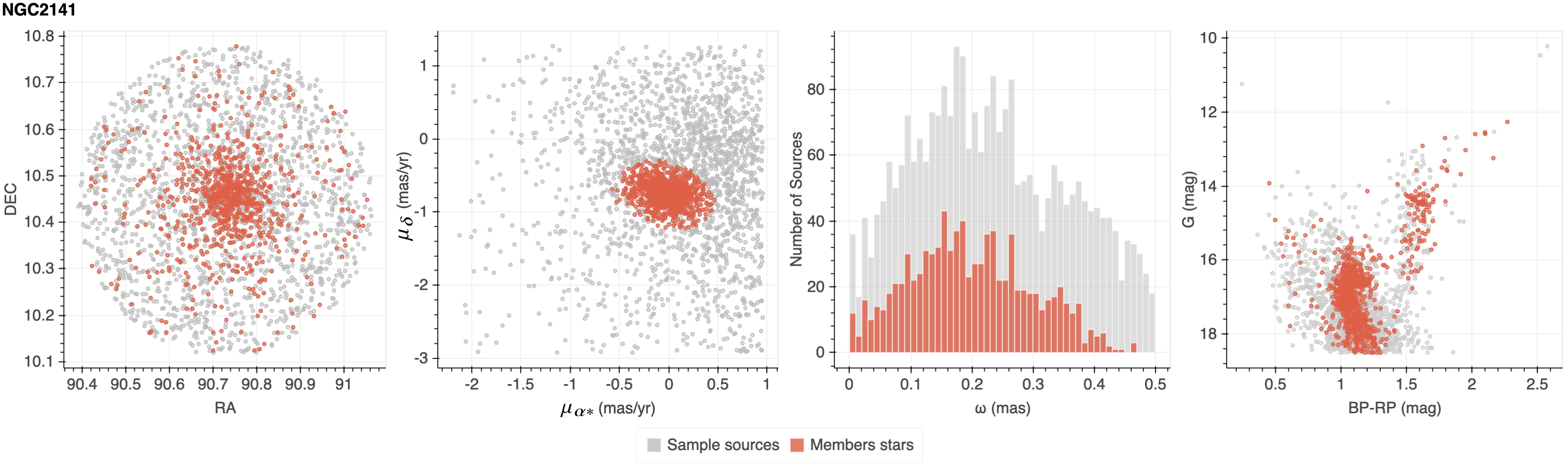}}\\
    \subfloat{\includegraphics[width=\textwidth]{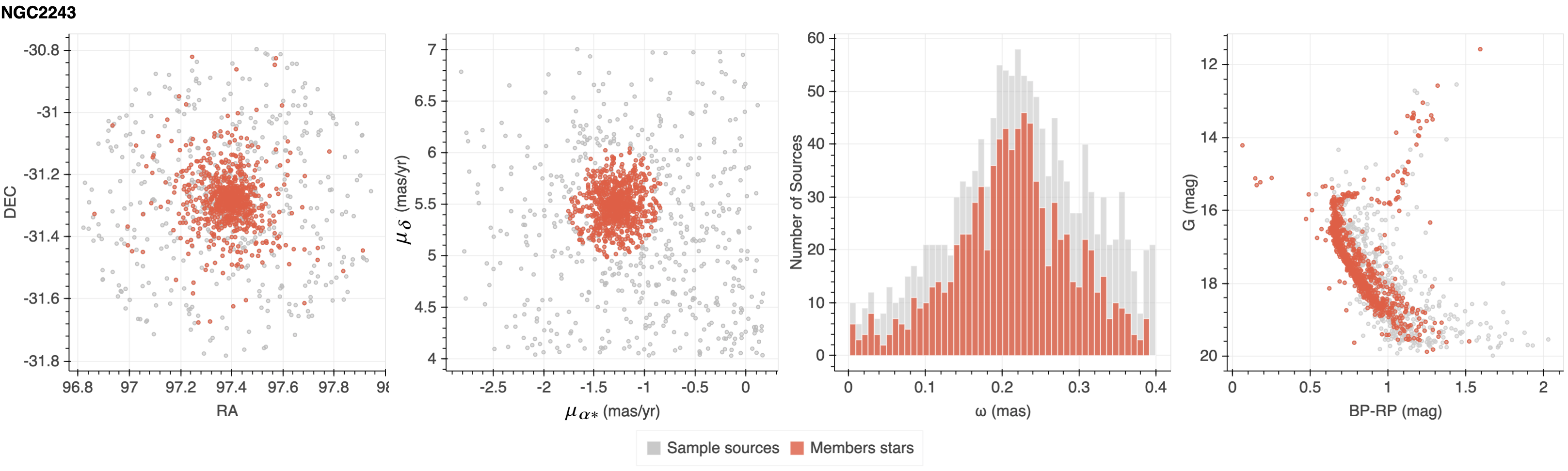}}\\
    \subfloat{\includegraphics[width=\textwidth]{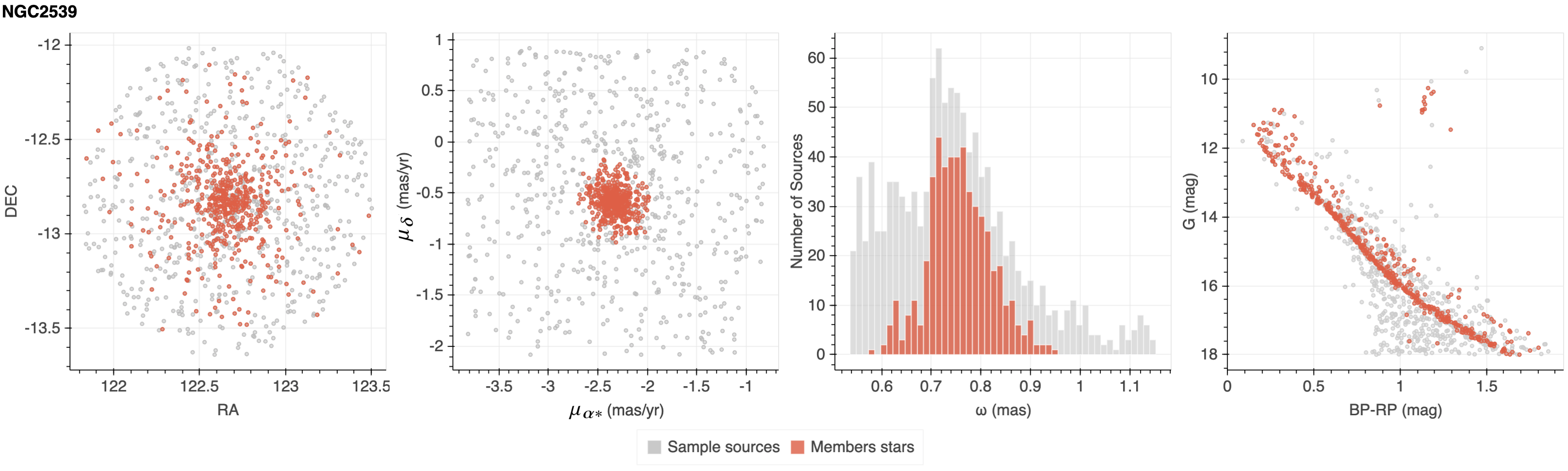}}
    \caption{The spatial distribution, the proper motion distribution, the parallax distribution and the CMD of the \textit{Sample sources} and the members identified by ML-MOC.}
    \label{fig:results}
\end{figure*}

\begin{figure*}\ContinuedFloat
    \centering
    \subfloat{\includegraphics[width=\textwidth]{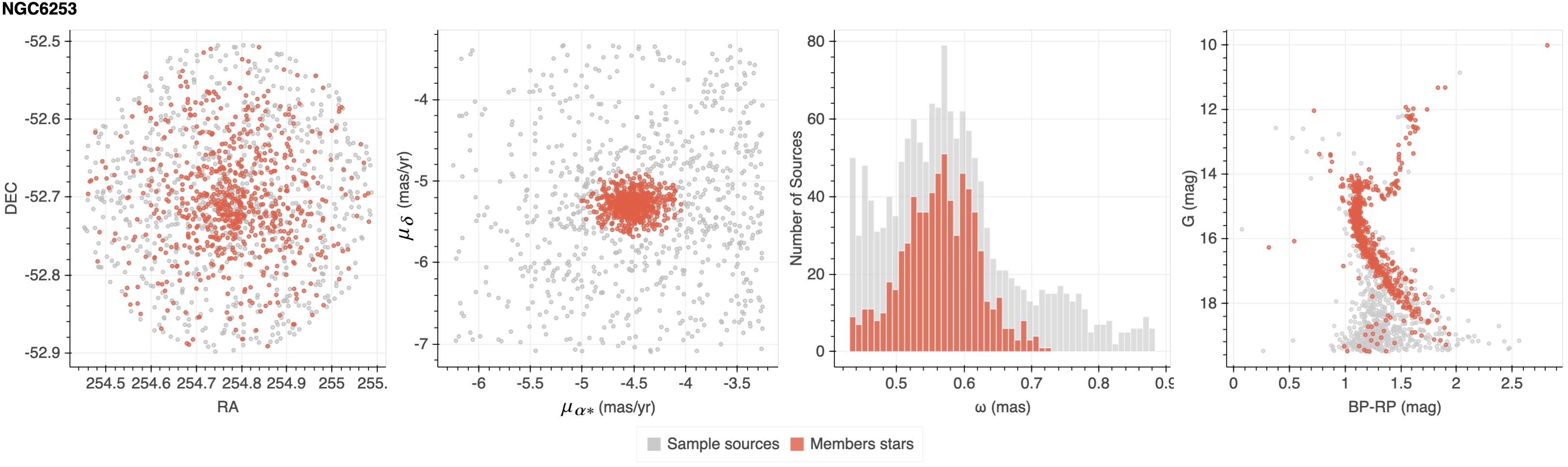}}\\
    \subfloat{\includegraphics[width=\textwidth]{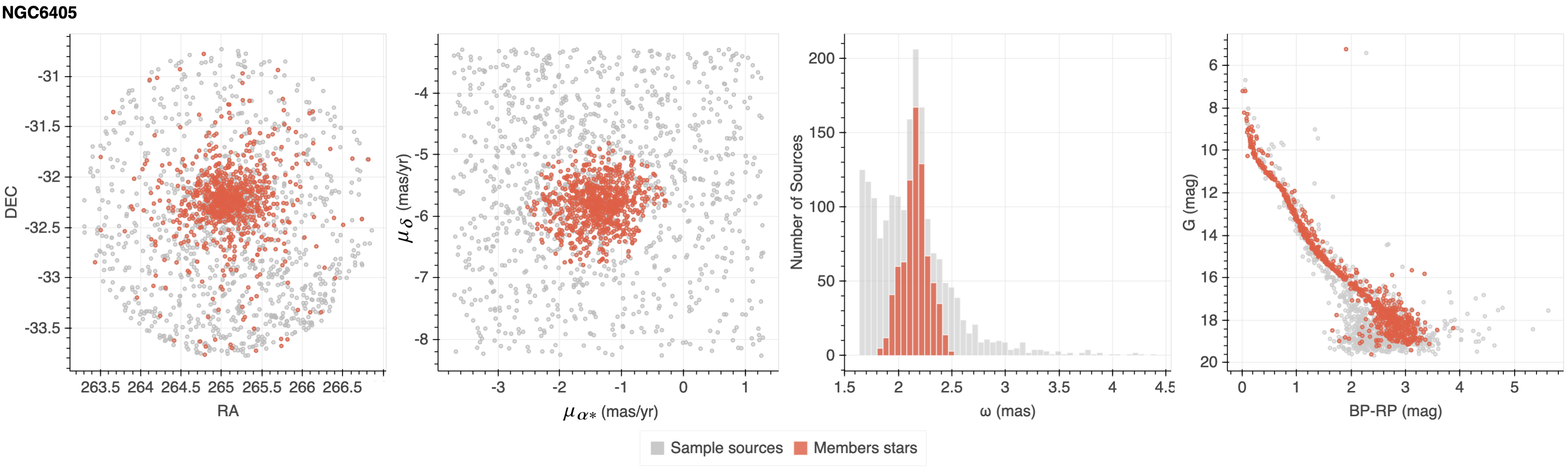}}\\
    \subfloat{\includegraphics[width=\textwidth]{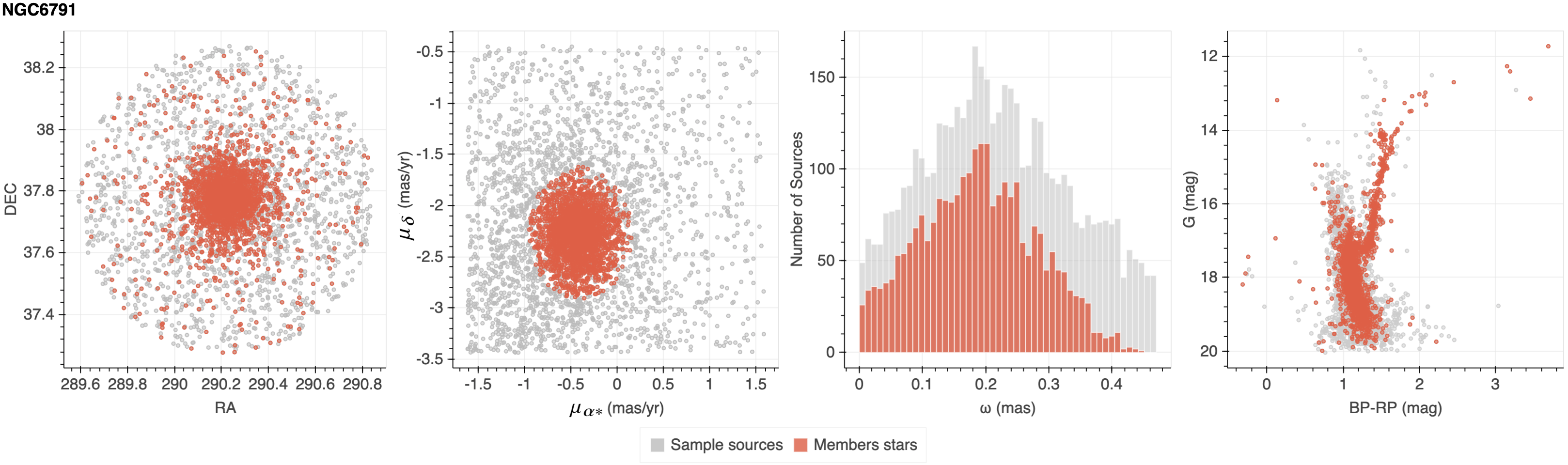}}\\
    \subfloat{\includegraphics[width=\textwidth]{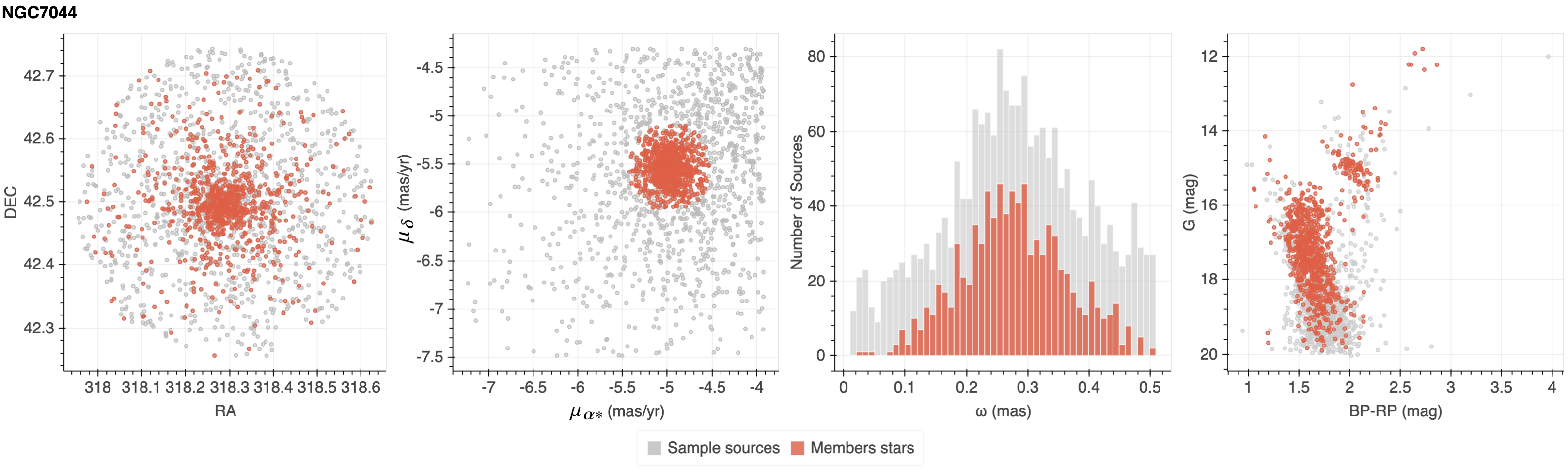}}
    \caption{The spatial distribution, the proper motion distribution, the parallax distribution and the CMD of the \textit{Sample sources} and the members identified by ML-MOC.}
\end{figure*}

\begin{figure*}\ContinuedFloat
    \centering
    \subfloat{\includegraphics[width=\textwidth]{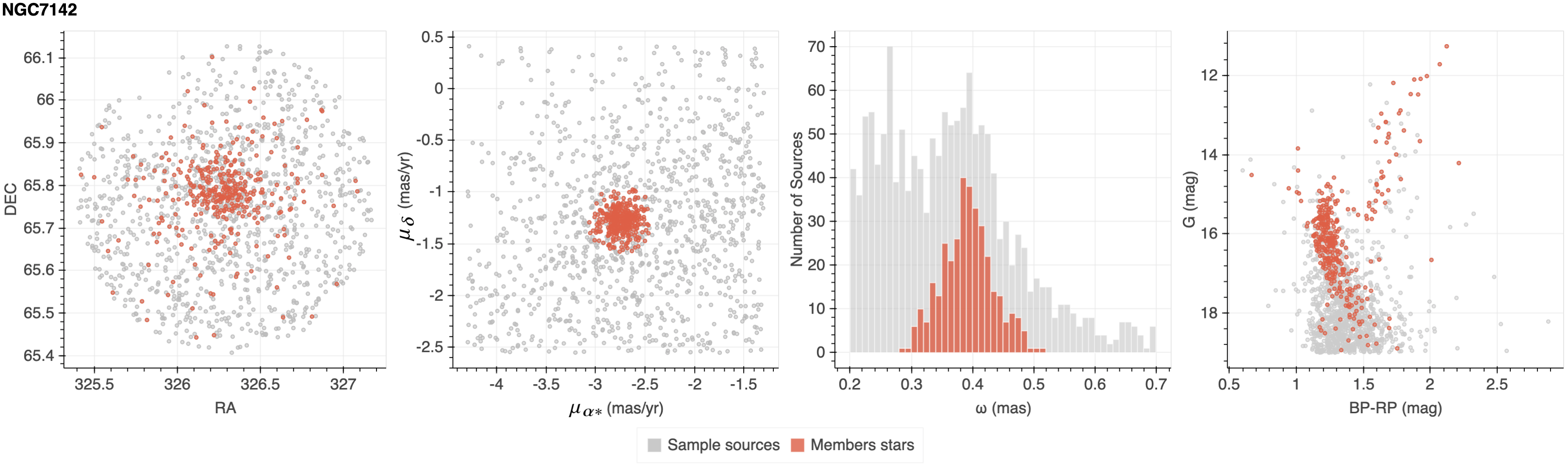}}\\
    \subfloat{\includegraphics[width=\textwidth]{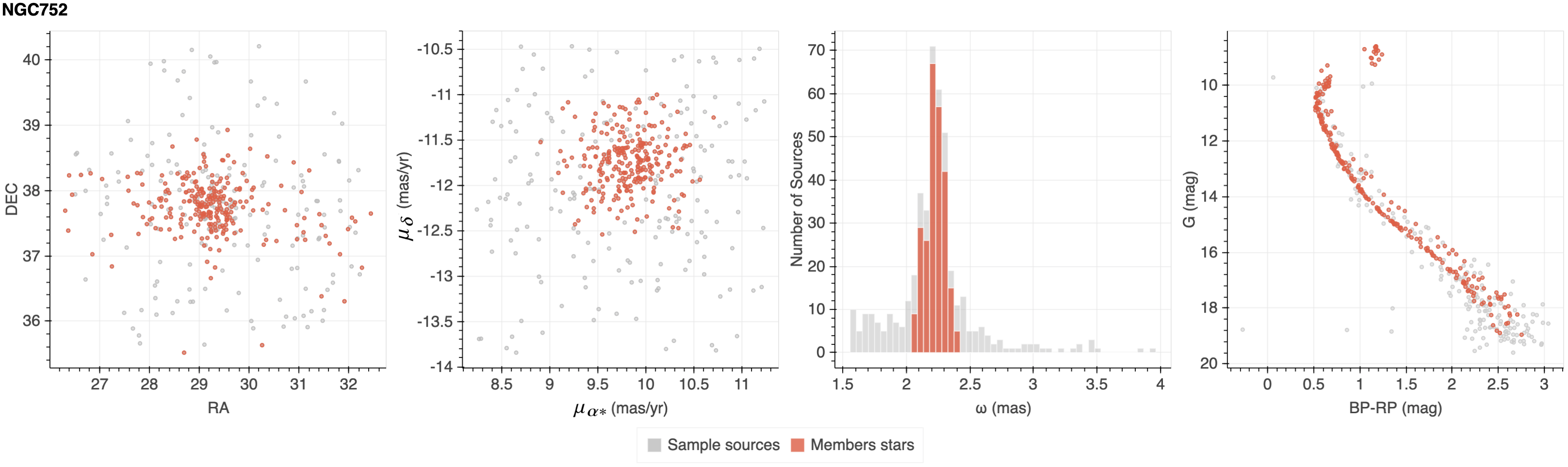}}\\
    \subfloat{\includegraphics[width=\textwidth]{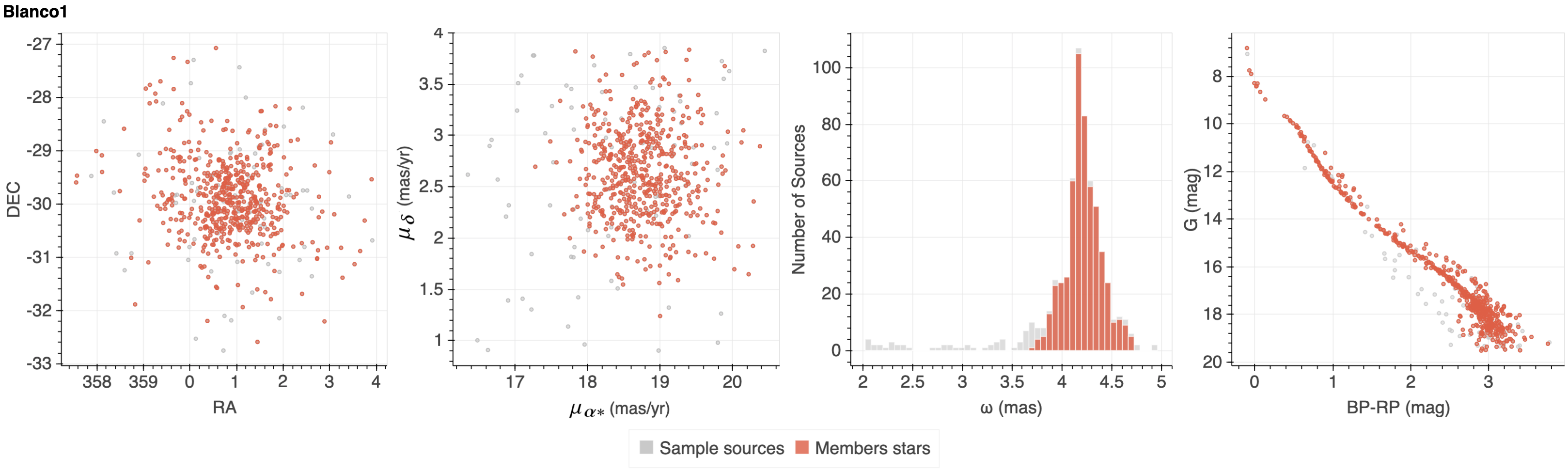}}\\
    \subfloat{\includegraphics[width=\textwidth]{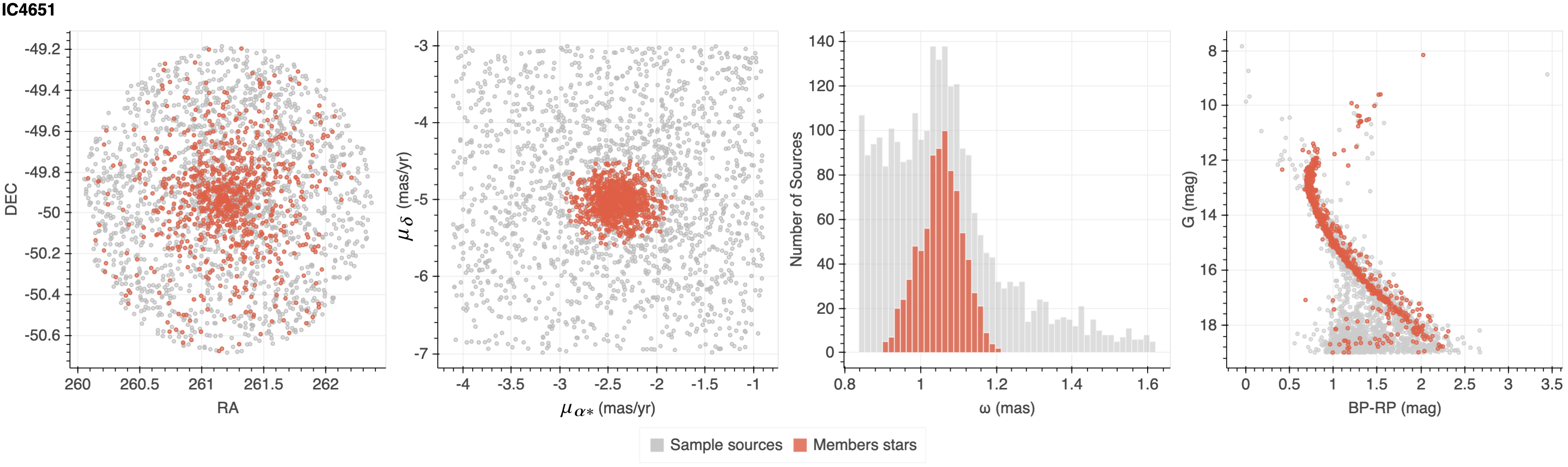}}
    \caption{The spatial distribution, the proper motion distribution, the parallax distribution and the CMD of the \textit{Sample sources} and the members identified by ML-MOC.}
\end{figure*}

\subsection{Degree of contamination}

To calculate an estimate for the degree of contamination, we download sources up to $r_{t2}$ i.e. the radius that encloses twice the area as enclosed by the $r_t$, and perform the algorithm as described in the Section ~\ref{sec:methodology} to identify sources satisfying the same criteria as members of the clusters but now between $r_t$ and $r_{t2}$. All the sources identified in the annular region of same area as the cluster area are considered as erroneously labelled members to calculate the degree of contamination. For M67, we find 1194 members stars within $r_t$ i.e. 98.686 arcmin and 23 additional sources up to $r_{t2}$ i.e. 139.563 arcmin. This gives a degree of contamination of 1.93\% for M67. In Figure~\ref{fig:M67_error}, we compare the member sources of M67 within the tidal radius with the sources in the annular region of same area as the cluster area. We observe that these 23 sources between $r_t$ and $r_{t2}$ are indistinguishable from the members sources within $r_t$ in all the plots except in the spatial distribution plot.

In Table~\ref{tab:contamination}, we show the estimated values of $r_c$ and $r_t$, and the degree of contamination for all the open clusters studied in this work except NGC 752, Be 18, Blanco 1, IC 4651 and Hyades. For a few clusters, marked by '*' in the Table~\ref{tab:contamination}, on applying ML-MOC to the sources up to $r_{t2}$, it fails to  clearly segregate the cluster members from the field stars. This is because when considering sources up to $r_{t2}$, we are adding a large number of field sources to the \textit{Sample sources}, hence reducing the proportion of cluster to field sources, which makes the GMM unreliable. For these clusters, we take sources brighter than $G\sim$18 mag to estimate the degree of contamination. For NGC 752, Be 18, Blanco 1, IC 4651 and Hyades, due to their sparse spatial distribution, we are unable to calculate the degree of contamination even with $G=18$ mag limit.

\subsection{Comparison with other clustering algorithms}

We compare the members extracted in this work using ML-MOC with the members identified by CG18 and Gao (\citealt{GaoM67} and \citealt{GaoNGC6405}, hereafter collectively referred as Gao18). CG18 employed the membership assignment code, Unsupervised Photometric Membership Assignment in Stellar Clusters (UPMASK), by \cite{UPMASK} on the proper motions and parallaxes from Gaia DR2. They consider sources only with $G$ < 18 mag and those that are located within a radius twice as large as the diameter reported by \cite{Dias2002}. Gao18 use a combination of GMM and Random Forest method on the astrometric and photometric measurements from Gaia DR2 to investigate the membership of open clusters M67 and NGC 6405. In Figure~\ref{fig:M67_comparison}, we show a comparison of the CMDs, proper motion and parallax distributions, of our member sources with those of CG18 and Gao18 for the cluster M67. The cluster members determined by all the three result in clean CMDs but this work and Gao18 find the member sources up to $G\sim$20 mag, whereas CG18 only goes up to $G\sim$18 mag.

\input{tables/results}

\subsection{Physical parameters of the cluster sample}
\label{subsec: stellar parameters}

We applied our method to twelve more open clusters located between $\sim$450 pc and $\sim$5500 pc. In Figure~\ref{fig:results}, we show the \textit{Sample sources} and the identified member sources for the rest of these open clusters, except for Berkeley 18 and Hyades which are shown in Figure~\ref{fig:Be18_result} and~\ref{fig:Hyades_result}, respectively. As mentioned in the Section ~\ref{sec:methodology}, it is difficult for the GMM to differentiate between field and cluster sources when there is a low concentration of cluster members and/or the parameter peaks coincide. In case of NGC 2539, our approach initially struggled at reliably identifying cluster members because the peak of $\mu_{\alpha*}$ for field sources coincides with that of cluster sources and there is a low cluster to field sources ratio. This was easily overcome by only considering sources brighter than $G =$ 18 mag, which improves the cluster to field sources ratio. Similarly for IC 4651, NGC 2141 and Berkeley 18, which all have overlapping peaks with the field sources, we consider sources brighter than $G\sim$19 mag to reliably extract cluster members. The figures comparing our identified members with CG18 and Gao18 are shown in Appendix~\ref{app:figures}.

To accurately compute the astrometric parameters of the cluster, we should consider only the most reliable member sources. Therefore, we use the sources that have a membership probability greater than 0.6 and lie within the estimated cluster radius. To reduce the uncertainty in the calculated parameter values, we only consider sources that are brighter than $G =$ 18 mag. In this work, we compute the mean proper motion and mean parallax of the cluster by taking a simple average of the individual proper motions and parallaxes of the most reliable member stars, without any consideration about their errors and covariances. For M67, the mean proper motion is determined to be ($\mu_{\alpha*}, \mu_\delta$) = ($10.981 \pm 0.006$, $-2.949 \pm 0.006$) mas/yr and mean parallax ($\omega$) = $1.135 \pm 0.002$ mas, using 965 bright and high probability members. Since reliable distances to the Gaia DR2 sources cannot be obtained by simply inverting the parallax, we take the mean of the individual
source distances obtained by \cite{BailerJones}. For M67, the mean distance is determined to be 860.7 pc. The centre coordinates of the clusters are computed using the Mean shift algorithm \citep{Meanshift}. We compare the astrometric parameters determined by us with CG18 in Table~\ref{tab:results}.

%% file: tables/contamination.tex
\begin{table*}
	\caption{An estimate of the degree of contamination in the member sources extracted by our algorithm.  Column 2 gives the cluster radius, columns 3 and 4 give core and tidal radii, respectively, as estimated from the King's profile fitting, column 5 gives the radius that encloses twice the area as enclosed by the tidal radius, column 6 gives the number of member stars for each cluster, column 7 gives the number of sources identified by the algorithm between the tidal radius of the cluster and the radius that encloses twice the area as the tidal radius, column 8 gives the degree of contamination for each cluster.}
	\label{tab:contamination}
	\begin{tabular}{l|ccccccc}
	\hline
        Clusters &r &$r_c$ &$r_t$ &$r_{t2}$ &N till $r_t$ &N between $r_t$ and $r_{t2}$ &Degree of contamination \\
        & (arcmin) & (arcmin) & (arcmin) & (arcmin) & & &($\%$)\\
    \hline
        M67 &53 &5.603 &98.686 &139.563 &1194 &23 &1.93 \\
        NGC2099 &41 &5.249 &57.471 &81.276 &1640 &103 &6.28 \\
        NGC2141* &11 &3.811 &16.883 &23.876 &828 &102 &12.32 \\
        NGC2243* &14 &1.207 &32.911 &46.543 &583 &12 &2.06 \\
        NGC2539* &23 &5.808 &39.508 &55.873 &466 &38 &8.15 \\
        NGC6253* &12 &3.364 &23.651 &33.448 &743 &47 &6.33 \\
        NGC6405 &53 &16.511 &58.869 &83.253 &688 &39 &5.67 \\
        NGC6791 &14 &3.041 &20.507 &29.001 &2422 &134 &5.53 \\
        NGC7044* &9 &1.597 &16.982 &24.016 &693 &46 &6.64 \\
        NGC7142* &12 &2.842 &22.719 &32.130 &316 &21 &6.65 \\
    \hline
\end{tabular}
\end{table*}

%% file: tables/results.tex
\begin{table*}
	\caption{The comparison of the astrometric parameters estimated in this work with CG18. The RA and DEC are expressed in degree, the mean proper motions along with the uncertainties are expressed in mas/yr, the mean parallaxes and the corresponding uncertainties are expressed in mas, and the distances are expressed in pc.}
	\label{tab:results}
	\begin{tabular}{l|cccccc|cccccc}
	\hline
        &\multicolumn{6}{c}{This work} &\multicolumn{6}{c}{CG18} \\
        Clusters &RA &DEC &$\mu_{\alpha*}$ &$\mu_\delta$ &$\omega$ &distance &RA &DEC &$\mu_{\alpha*}$ &$\mu_\delta$ &$\omega$ &distance \\
        & & &$\epsilon_{\mu_{\alpha*}}$ &$\epsilon_{\mu_\delta}$ &$\epsilon_\omega$ & & & &$\epsilon_{\mu_{\alpha*}}$ &$\epsilon_{\mu_\delta}$ &$\epsilon_\omega$ & \\
    \hline
        M67 &132.852 &11.836 &-10.981 &-2.949 &1.135 &860.7 &132.846 &11.814 &-10.986 &-2.964 &1.135 &859.1 \\
        & & &0.006 &0.006 &0.002 & & & &0.008 &0.008 &0.002 & \\
        NGC2099 &88.064 &32.547 &1.928 &-5.636 &0.663 &1466.9 &88.074 &32.545 &1.924 &-5.648 &0.666 &1438.1 \\
        & & &0.004 &0.004 &0.002 & & & &0.006 &0.005 &0.002 & \\
        NGC2141 &90.742 &10.455 &-0.025 &-0.750 &0.197 &3812.3 &90.734 &10.451 &-0.028 &-0.767 &0.196 &4441.3 \\
        & & &0.006 &0.005 &0.004 & & & &0.006 &0.007 &0.004 & \\
        NGC2243 &97.390 &-31.283 &-1.285 &5.489 &0.213 &3606.0 &97.395 &-31.282 &-1.279 &5.488 &0.211 &4167.8 \\
        & & &0.005 &0.006 &0.003 & & & &0.006 &0.006 &0.003 & \\
        NGC2539 &122.670 &-12.845 &-2.337 &-0.584 &0.757 &1280.5 &122.658 &-12.834 &-2.331 &-0.584 &0.754 &1277.4 \\
        & & &0.006 &0.006 &0.003 & & & &0.006 &0.007 &0.003 & \\
        NGC6253 &254.769 &-52.713 &-4.521 &-5.289 &0.567 &1718.6 &254.778 &-52.712 &-4.537 &-5.280 &0.563 &1689.7 \\
        & & &0.005 &0.005 &0.002 & & & &0.009 &0.007 &0.003 & \\
        NGC6405 &265.088 &-32.285 &-1.360 &-5.816 &2.157 &460.6 &265.069 &-32.242 &-1.306 &-5.847 &2.172 &454.3 \\
        & & &0.015 &0.013 &0.005 & & & &0.018 &0.016 &0.004 & \\
        NGC6791 &290.226 &37.776 &-0.422 &-2.2745 &0.186 &4086.6 &290.221 &37.778 &-0.421 &-2.269 &0.192 &4530.8 \\
        & & &0.0039 &0.0048 &0.002 & & & &0.005 &0.006 &0.002 & \\
        NGC7044 &318.287 &42.494 &-4.976 &-5.523 &0.273 &3273.5 &318.284 &42.494 &-4.976 &-5.526 &0.273 &3315.6 \\
        & & &0.005 &0.006 &0.003 & & & &0.007 &0.007 &0.003 & \\
        NGC7142 &326.287 &65.753 &-2.743 &-1.283 &0.392 &2401.3 &326.290 &65.782 &-2.747 &-1.288 &0.392 &2376.4 \\
        & & &0.006 &0.006 &0.002 & & & &0.008 &0.007 &0.003 & \\
        NGC752 &29.156 &37.809 &9.825 &-11.724 &2.229 &443.8 &29.223 &37.794 &9.810 &-11.713 &2.239 &441.0 \\
        & & &0.018 &0.020 &0.005 & & & &0.019 &0.019 &0.005 & \\
        Blanco1 &1.030 &-29.926 &18.724 &2.644 &4.207 &236.6 &0.853 &-29.958 &18.739 &2.602 &4.210 &235.9 \\
        & & &0.023 &0.021 &0.008 & & & &0.023 &0.022 &0.007 & \\
        Be18 &80.472 &45.396 &0.843 &-0.078 &0.139 &4652.0 &80.531 &45.442 &0.849 &-0.057 &0.152 &5523.5 \\
        & & &0.012 &0.009 &0.004 & & & &0.010 &0.008 &0.005 & \\
        IC4651 &261.230 &-49.873 &-2.424 &-5.039 &1.055 &933.0 &261.212 &-49.917 &-2.410 &-5.064 &1.056 &921.3 \\
        & & &0.009 &0.010 &0.003 & & & &0.008 &0.008 &0.002 & \\
        Hyades &66.824 &16.267 &107.076 &-25.082 &21.273 &47.1 & & & & & & \\
        & & &0.656 &0.556 &0.089 & & & & & & & \\
    \hline
\end{tabular}
\end{table*}

%% file: chapters/conclusions.tex
\section{Discussion}
\label{sec:conclusions}

In this work, we presented, ML-MOC, a new method for open cluster membership determination using only astrometric measurements from Gaia DR2 and no a priori information about the cluster parameters. We employed a combination of well-known Machine Learning algorithms, i.e. k-Nearest Neighbours algorithm and Gaussian Mixture Model, to estimate the membership probability of individual sources down to $G\sim$20 mag. Our approach does not rely on any strong physical assumption concerning the nature of the cluster (no assumptions on the density profile modelling or on the structure in the photometric space). We applied ML-MOC to fifteen open clusters that cover a wide parameter space in terms of their distances, $\sim$47 pc to $\sim$5500 pc, ages, 0.53 Gyr to 8.89 Gyr, metallicities [Fe/H], $-$0.54 to 0.43, and extinctions, 0.2 mag to 1.7 mag. The cluster members identified by ML-MOC successfully produce clean CMDs. On considering the radial velocity verified members by WOCS for M67, NGC 6791 and NGC 6253 as ground truth, ML-MOC retrieves more members than CG18 while maintaining a similar number of false classifications.

The cluster centres, mean proper motions and mean parallaxes of the clusters measured with ML-MOC are in excellent agreement, as shown in Table~\ref{tab:results}, with those determined by CG18 who make use of the same input catalogue (Gaia DR2), while identifying members at greater photometric depth and including nearby clusters like Hyades.

The open clusters whose proper motions and parallaxes merge with the field sources are challenging for ML-MOC. Most of these clusters can be resolved by considering the brighter sources and reducing the initial selection radius. However, we cannot entirely exclude the possibility that some of the sources identified as member stars in such clusters are field stars. To estimate the field stars erroneously labelled as cluster members we evaluate the degree of contamination. For M67 and NGC 2243, the contamination is estimated at just 2$\%$. In six of the studied clusters, we determine the contamination between 5$\%$ and 7$\%$. The highest degree of contamination was estimated in NGC 2539 and NGC 2141 at 8.15$\%$ and 12.32$\%$ respectively. The high contamination is likely due to the overlap in the peak of field sources and cluster members in ($\mu_{\alpha*}$) for NGC 2539 and ($\mu_{\alpha*}$, $\omega$) for NGC 2141. 

ML-MOC is a reliable and robust approach to extract open cluster members when there is a distinctive separation of the members stars from the field sources in the ($\mu_{\alpha*}$,  $\mu_\delta$, $\omega$) space. The non-reliance of ML-MOC on spatial information will help us to find non-circular morphology and study the internal dynamical processes including those that influence the formation of tidal tails \citep{Souradeep17,tails2019}. As an example, in the spatial distribution of the cluster NGC 752 in Figure~\ref{fig:results}, the presence of possible tidal tails can be seen. The comprehensive information about the membership of a large number of open clusters will enable us to study a host of issues of relevance in Astrophysics such as, the dynamical evolution \citep{evaporation2013,DynamicEvolution2019}, signatures of mass segregation \citep{massSeg2009,massSeg_luminosity2020}, evaporation of lower-mass cluster members \citep{evaporation2013}, and dissolving of old open clusters \citep{dissolution,DynamicEvolution2019}, and studying the luminosity functions \citep{luminosity2018, massSeg_luminosity2020},  initial mass functions \citep{IMF_Binary1997,IMF2016}, and estimating primordial binary population \citep{IMF_Binary1997, Binary2007}, in young open clusters. The algorithm also has a great potential to allow the study of exotic stellar populations such as the blue straggler stars in open clusters \citep{Bhattacharya19,Vaidya20,Rain2020}, and their formation mechanisms. Moreover, with the recent release of Gaia Early Data Release 3 \footnote{https://www.cosmos.esa.int/web/gaia/earlydr3}, and later Gaia Data Release 3 which will provide more precise astrometric, photometric, and radial velocity information, our algorithm will improve the reliability of membership determination and classify even dimmer sources.

%% file: appendices/Be18.tex
\section{Berkeley 18}
\label{app:Be18}

\begin{figure*}
	\includegraphics[width=\textwidth]{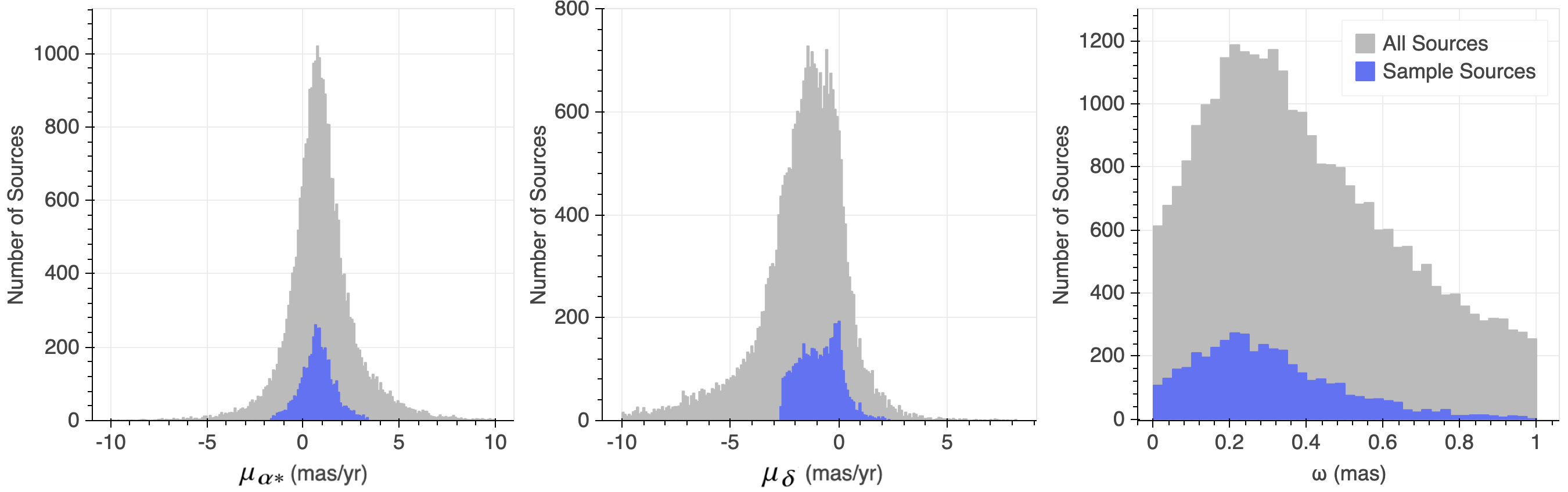}
    \caption{Distribution of Be18 \textit{All sources} and \textit{Sample sources} in $\mu_{\alpha*}, \mu_\delta, \omega$. The peak of $\mu_{\alpha*}$ and $\omega$ of field sources overlaps with the corresponding peaks of cluster members.}
    \label{fig:Be18_stage1}
\end{figure*}

\begin{figure*}
	\includegraphics[width=\textwidth]{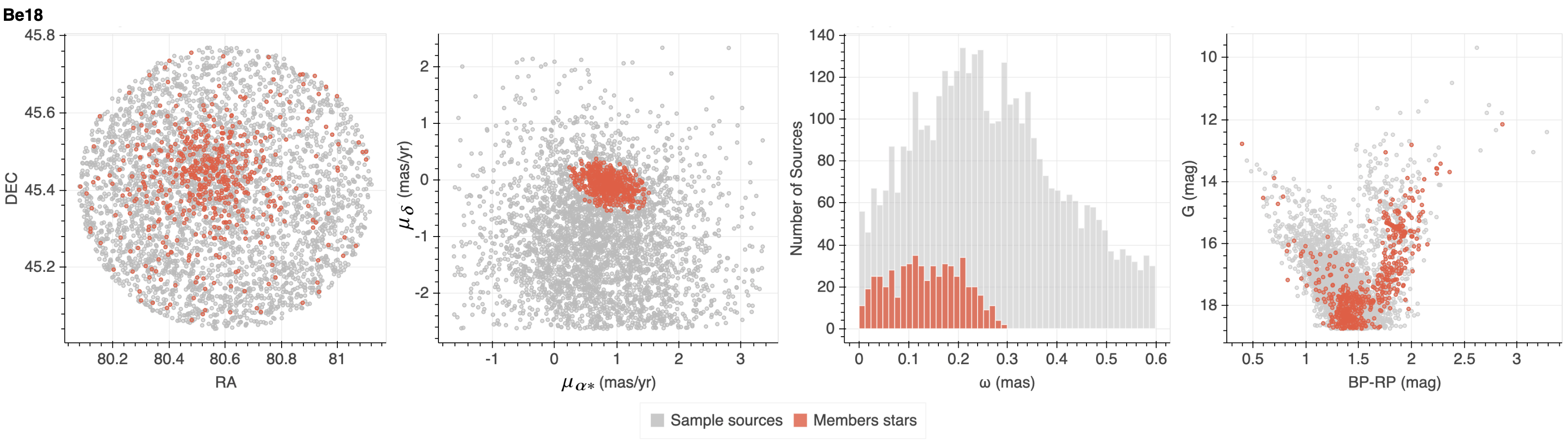}
    \caption{The grey points represent the \textit{Sample sources} and the red points are the identified cluster members for Be18 that are within 22 arcmin and brighter than $G$ = 18.75 mag.}
    \label{fig:Be18_result}
\end{figure*}

\begin{figure*}
	\includegraphics[width=\textwidth]{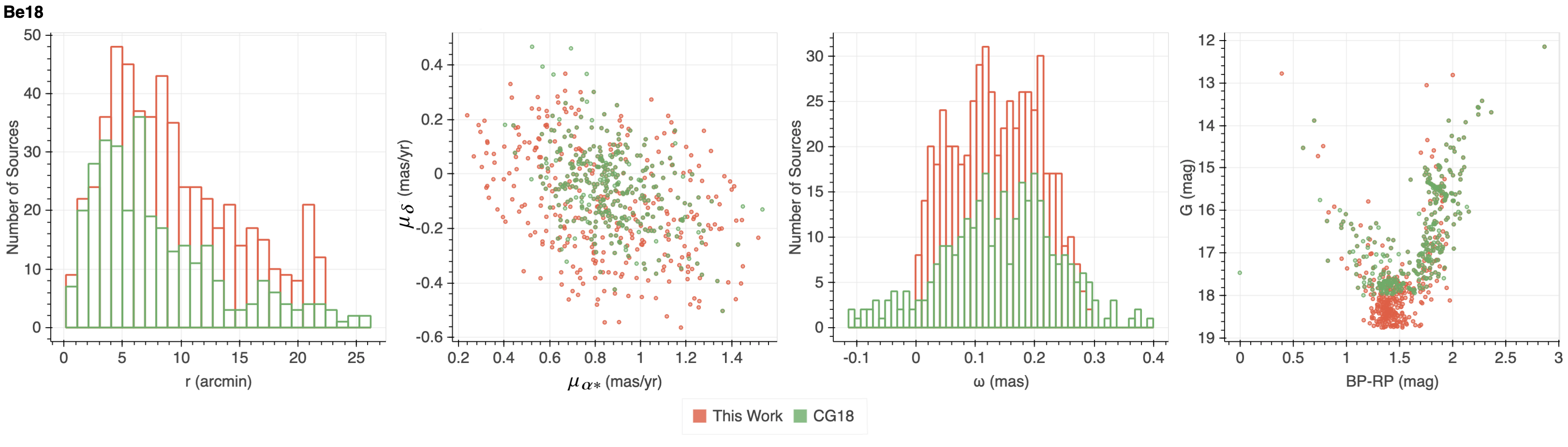}
    \caption{Comparing the cluster members of Be18 determined by ML-MOC with CG18.}
    \label{fig:Be18_comparison}
\end{figure*}

Berkeley 18 is an old open cluster with the estimated distance of $\sim$5.8 kpc \citep{Kaluzny}. On directly applying the method described in this work we fail to segregate the likely cluster members of Be18 from the field stars. This is because of (1) very low cluster members to field sources ratio and (2) the overlapping mean values of $\mu_{\alpha*}$ and $\omega$ for cluster and field sources (shown in Figure~\ref{fig:Be18_stage1}). These are clear violations of the criteria, as mention in second stage in section~\ref{sec:methodology}, required for reliable determination of cluster members by the Gaussian Mixture Model. To overcome this, we need to reduce the number of field sources while retaining the maximum  cluster members. We achieve this by only considering sources within 22 arcmin (according to K13 the tidal radius, $r_t$, is 26 arcmin and cluster radius is 16.5 arcmin) and brighter than $G$ = 18.75 mag. In Figure~\ref{fig:Be18_result} we show the result after applying these limits and in Figure~\ref{fig:Be18_comparison} we compare our extracted members with CG18. It is possible to reduce the radius at the expense of $G$-mag and vice-a-versa, so depending on the particular science case one can decide these limits. 

%% file: appendices/Hyades.tex
\section{Hyades}
\label{app:Hyades}

\begin{figure*}
	\includegraphics[width=\textwidth]{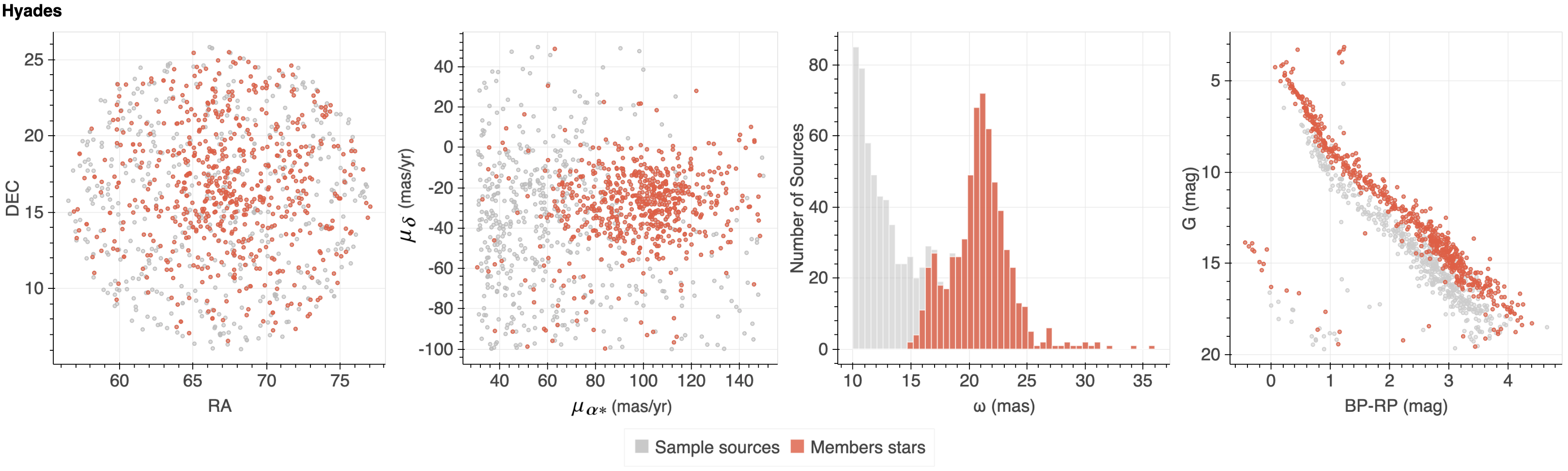}
    \caption{The grey points represent the \textit{Sample sources} and the red points are the identified cluster members for Hyades that are within 600 arcmin from the cluster centre.}
    \label{fig:Hyades_result}
\end{figure*}

Hyades (Melotte 25) is a young open cluster at an age of 625 $-$ 750 Myr \citep{Hyades_age1,Hyades_age2,Hyades_age3} and is located at an average distance of 46.75 $\pm$ 0.46 pc \citep{GaiaCollab_2017}. CG18 excluded it from their study because it has a very large extension across the sky which makes the membership determination difficult. We take the search radius of 10 degrees, centred on the cluster, to identify its member sources. Three other known open clusters, NGC 1647, NGC 1662 and UPK 385, are also located in this region of the sky. When the region has more than one cluster, ML-MOC identifies the members of the most denser cluster in the ($\mu_{\alpha*}$,  $\mu_\delta$, $\omega$) space. To identify the members of the other clusters, we can either remove the identified members of the first cluster before performing the algorithm again on the region or (if one knows a priori about the other cluster) we can manually restrict the initial search space. For Hyades, we restrict the search by only considering sources with parallax greater than 10 mas, to reduce the proportion of field stars and to avoid detection of other open clusters present in the search region. As Hyades is a very sparse cluster, we apply the kNN algorithm on all the considered sources. Figure~\ref{fig:Hyades_result} shows the extracted cluster members. WOCS \citep{Hyades_WOCS} identifies 54 sources as members of the Hyades cluster, in our searched region of the sky, of which we successfully retrieve 53 (98.14$\%$) sources.

%% file: appendices/additional_figures.tex
\section{Additional figures}
\label{app:figures}

\begin{figure*}
	\includegraphics[width=\textwidth]{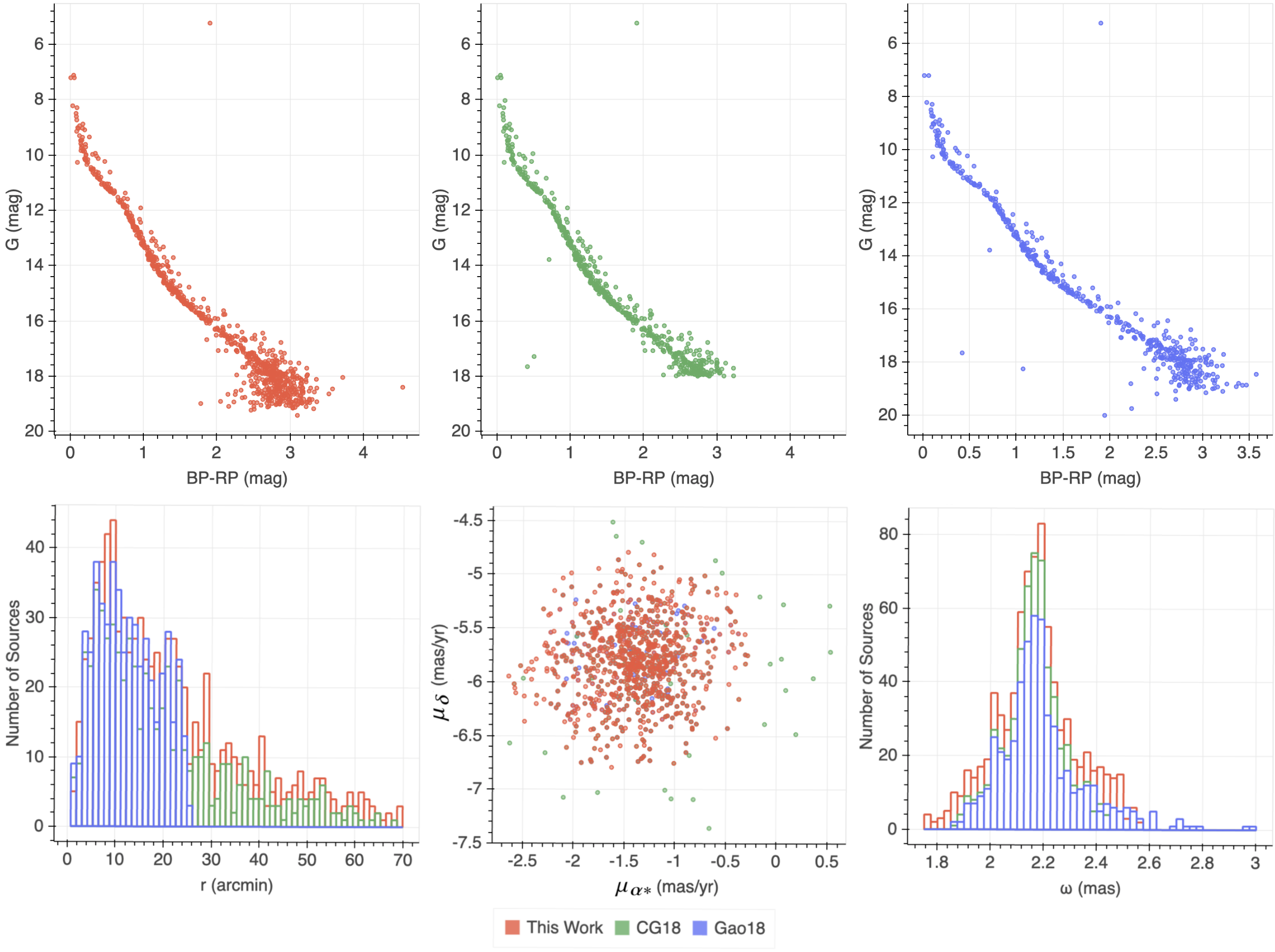}
    \caption{Upper Panel:  The CMDs of members identified for the cluster NGC 6405 by our algorithm, CG18 and Gao18.  Lower Panel: The radial distribution, the proper motion distribution, and the parallax distribution of members by the three algorithms.}
    \label{fig:NGC6405_comparison}
\end{figure*}

\begin{figure*}
    \centering
    \subfloat{\includegraphics[width=\textwidth]{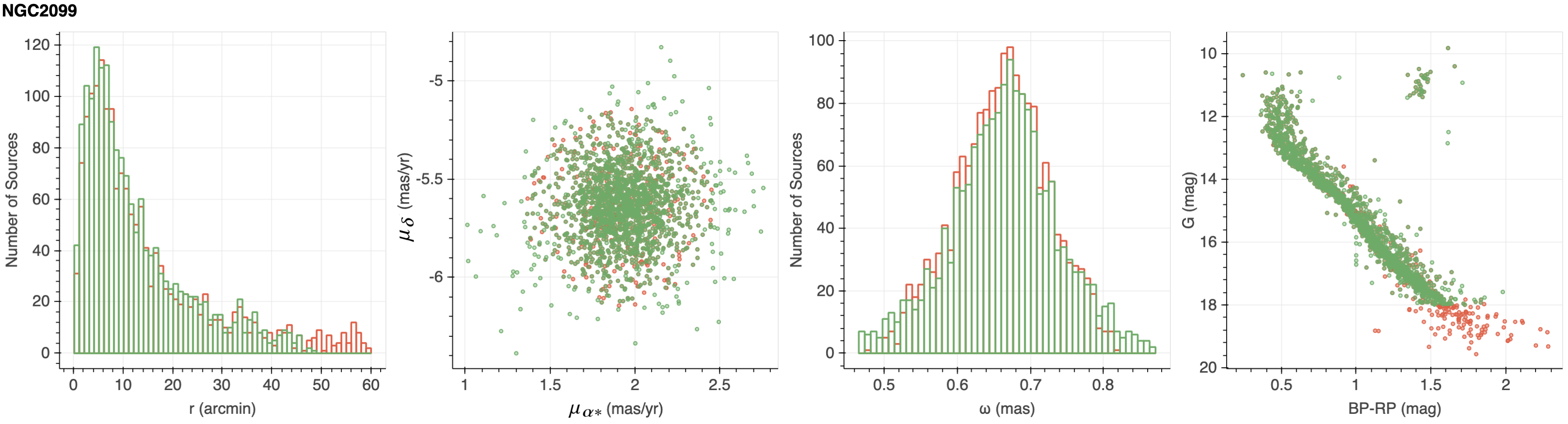}}\\
    \subfloat{\includegraphics[width=\textwidth]{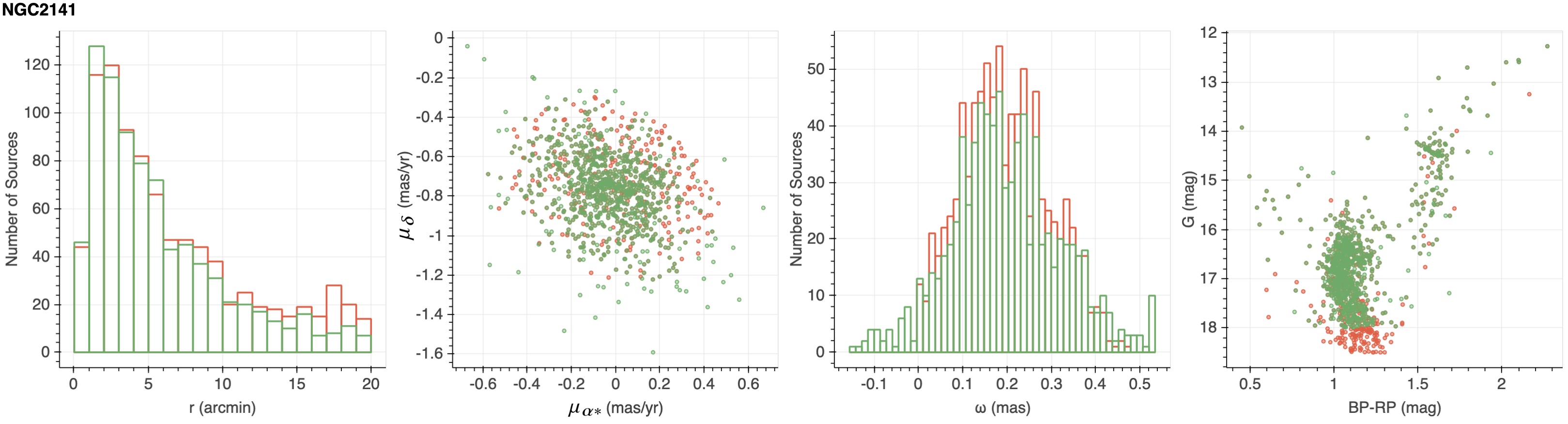}}\\
    \subfloat{\includegraphics[width=\textwidth]{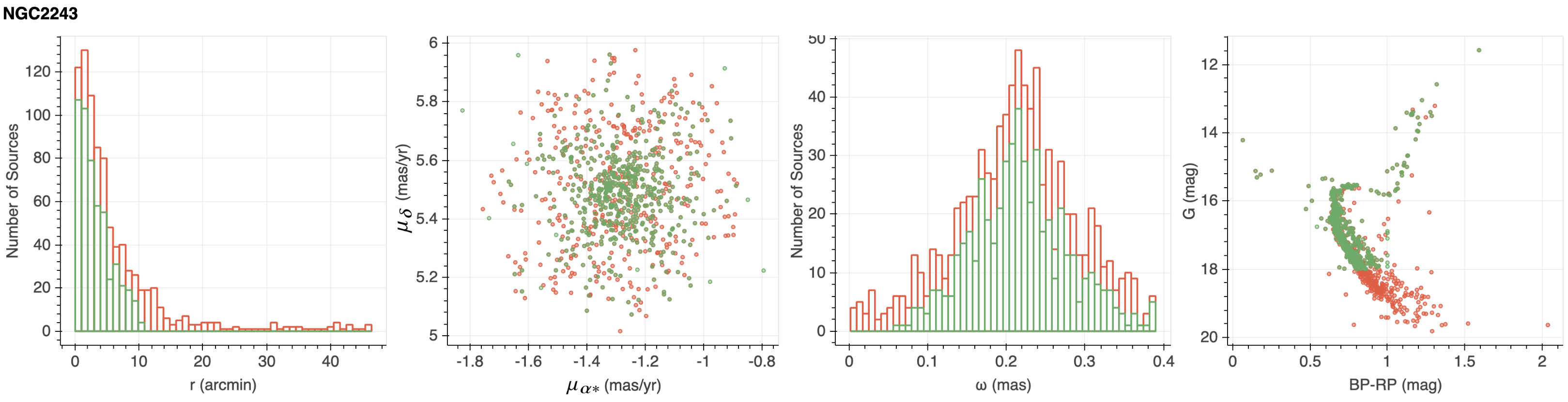}}\\
    \subfloat{\includegraphics[width=\textwidth]{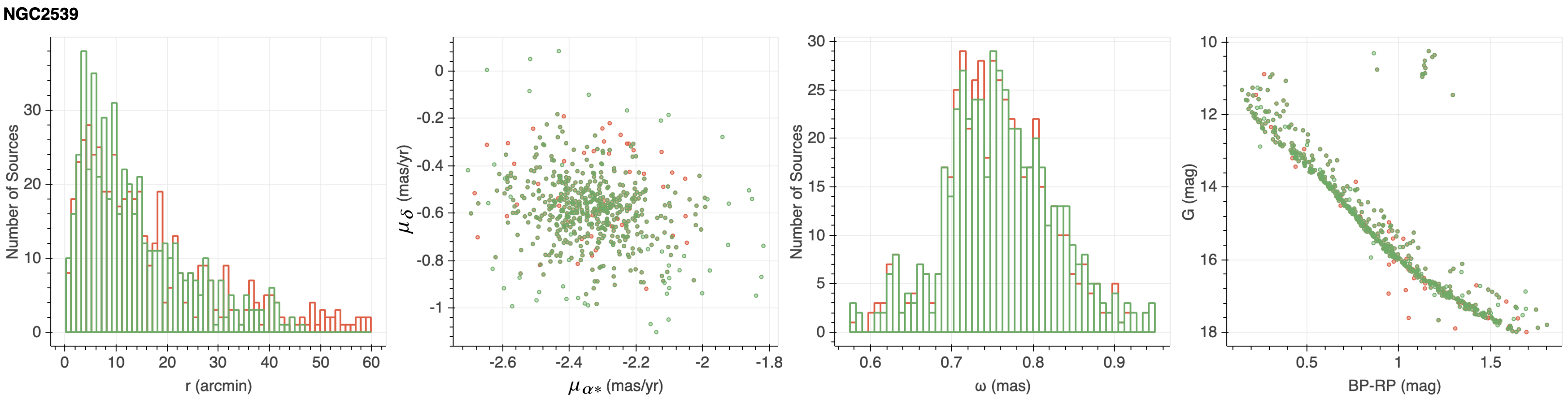}}
    \caption{Comparing the cluster member sources determined in this work with the members by CG18. The sources in red are extracted by ML-MOC. CG18 member sources (i.e. sources with membership probability greater than 0.5) are shown in green.}
    \label{fig:comparisons}
\end{figure*}

\begin{figure*}\ContinuedFloat
    \centering
    \subfloat{\includegraphics[width=\textwidth]{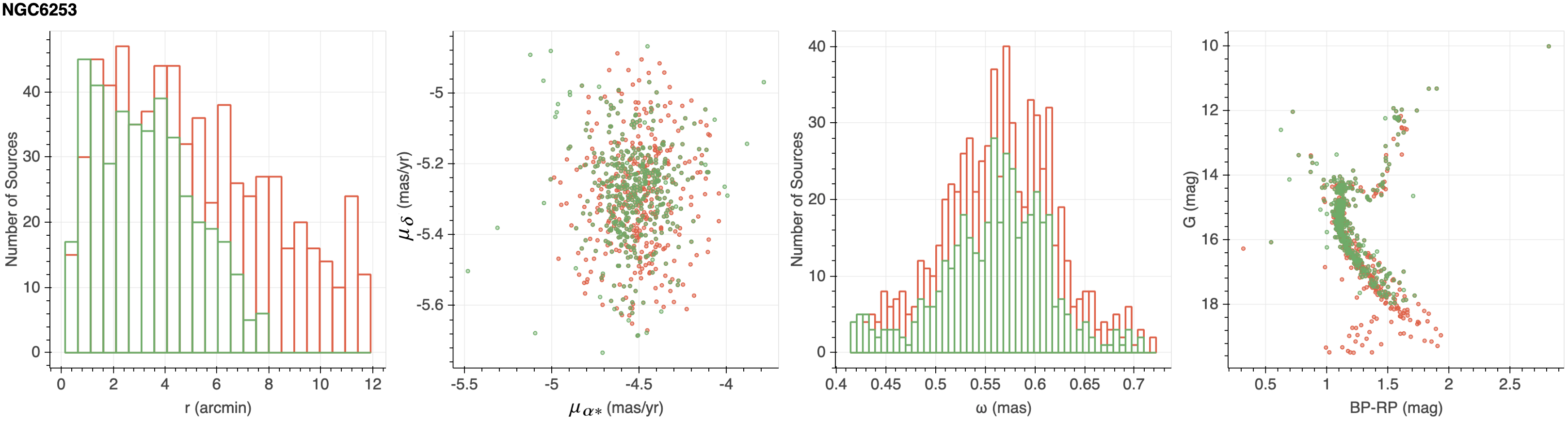}}\\
    \subfloat{\includegraphics[width=\textwidth]{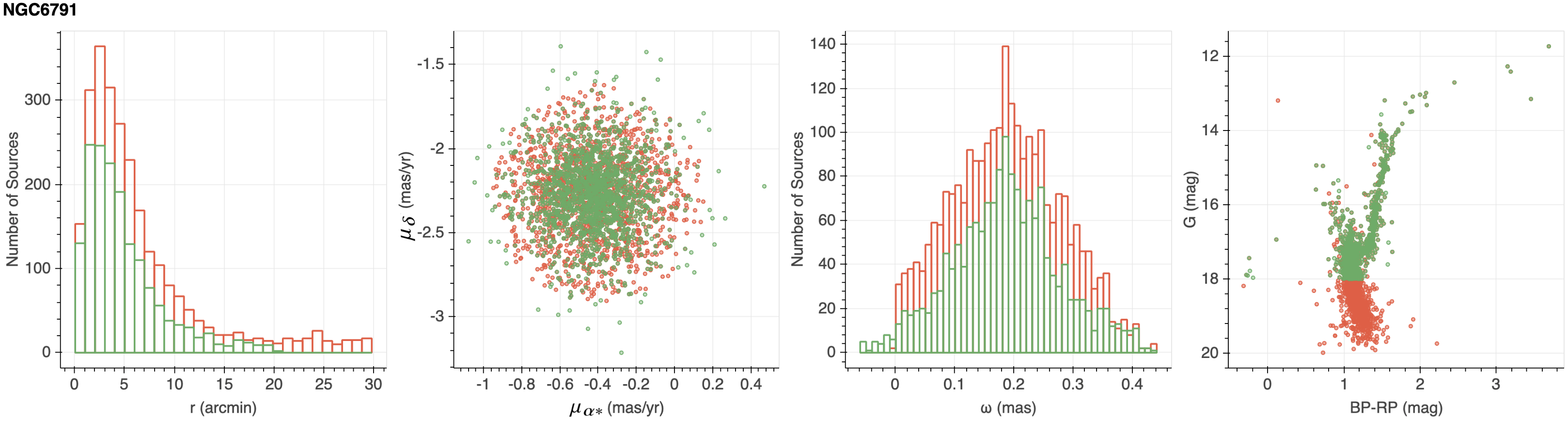}}\\
    \subfloat{\includegraphics[width=\textwidth]{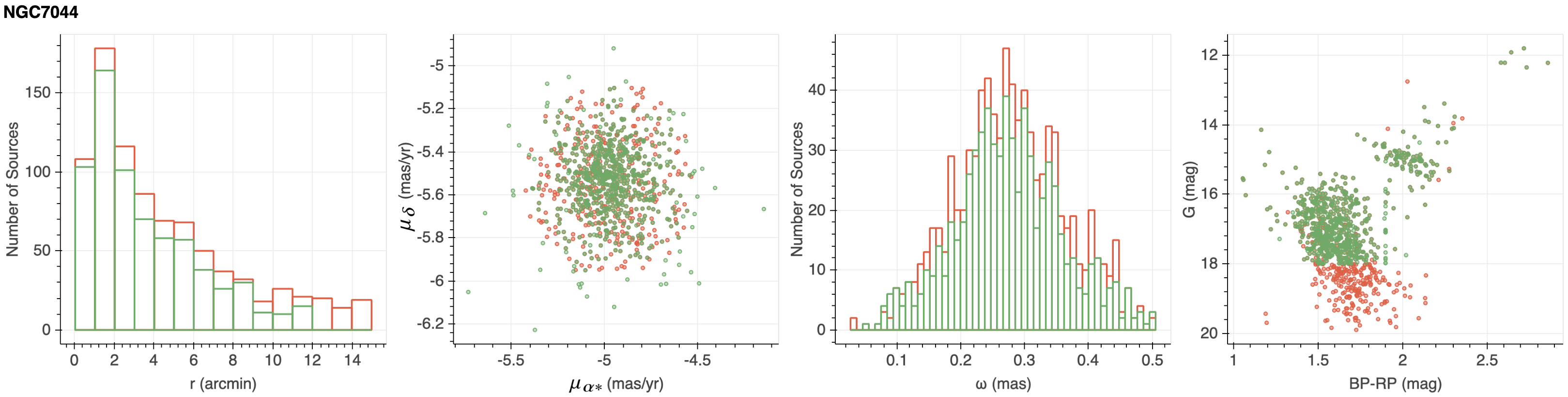}}\\
    \subfloat{\includegraphics[width=\textwidth]{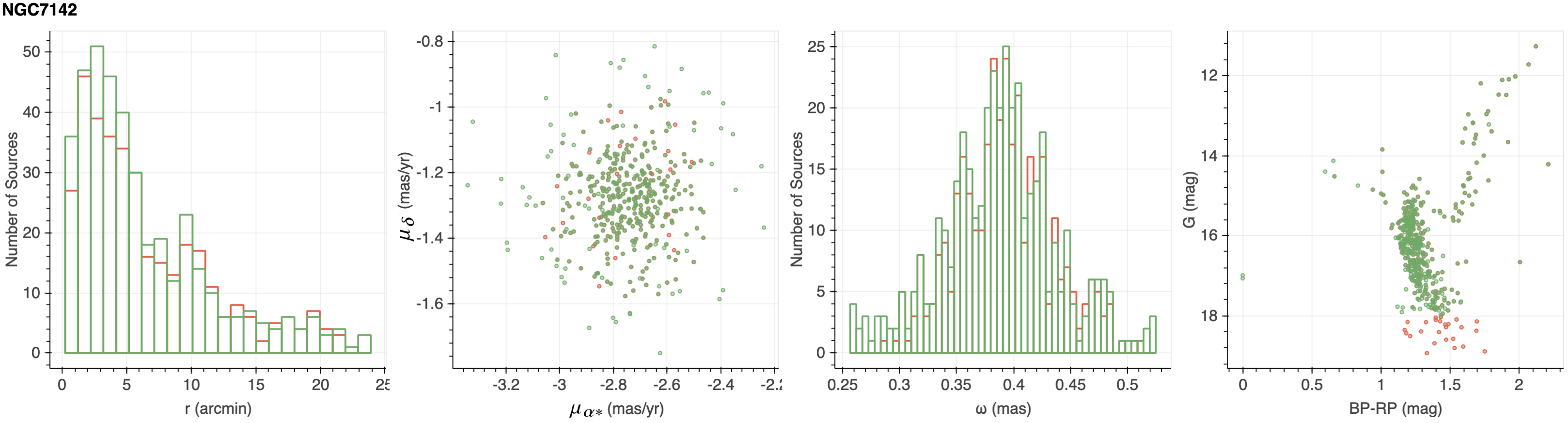}}
    \caption{Comparing the cluster member sources determined in this work with the members by CG18. The sources in red are extracted by ML-MOC. CG18 member sources (i.e. sources with membership probability greater than 0.5) are shown in green.}
\end{figure*}

\begin{figure*}\ContinuedFloat
    \centering
    \subfloat{\includegraphics[width=\textwidth]{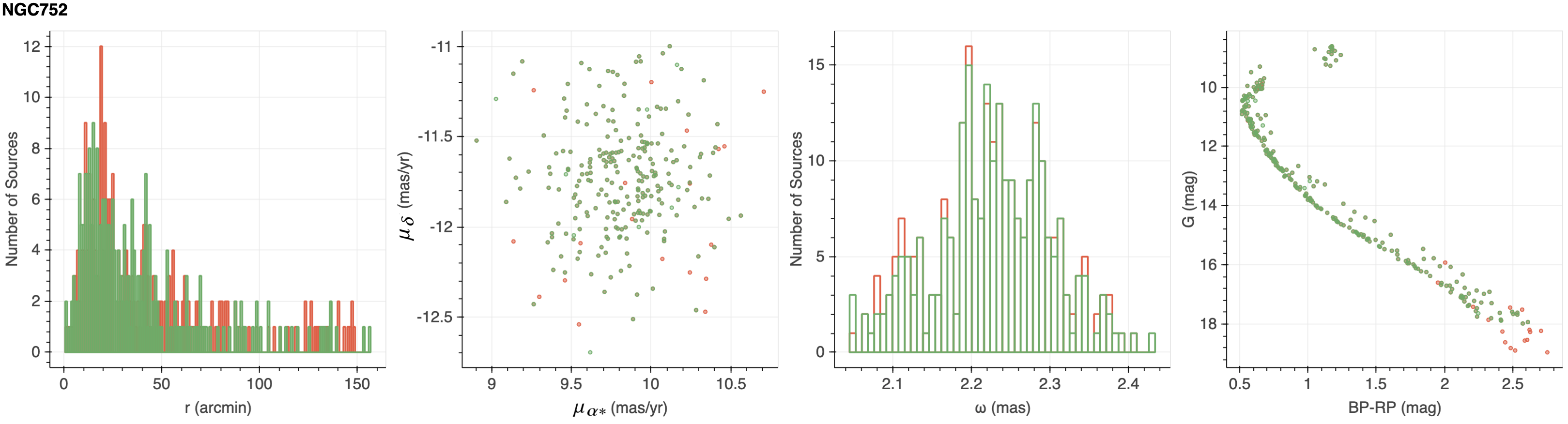}}\\
    \subfloat{\includegraphics[width=\textwidth]{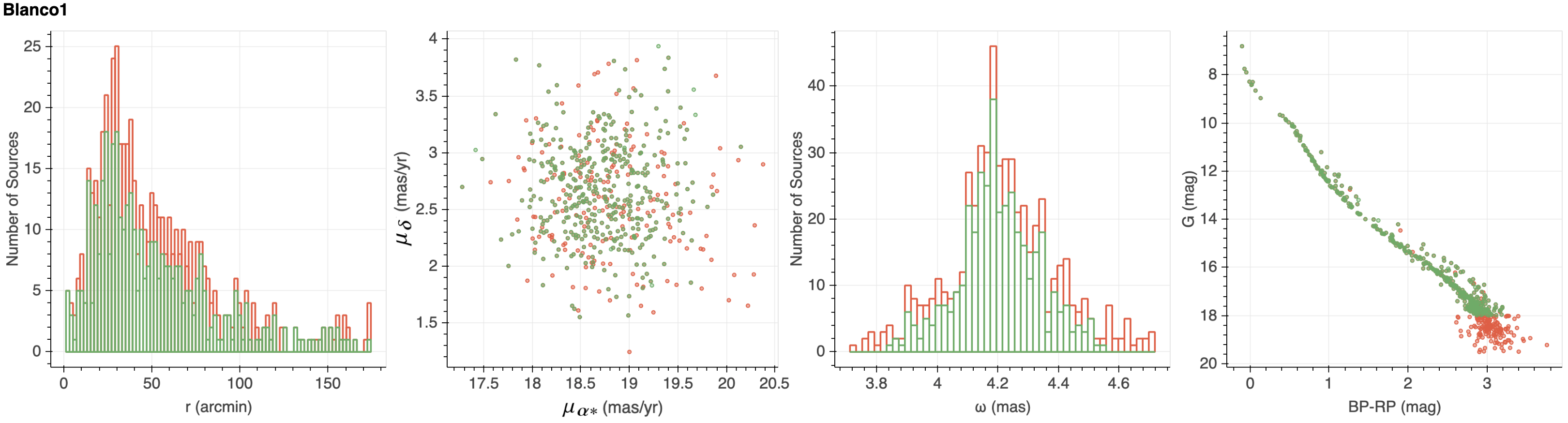}}\\
    \subfloat{\includegraphics[width=\textwidth]{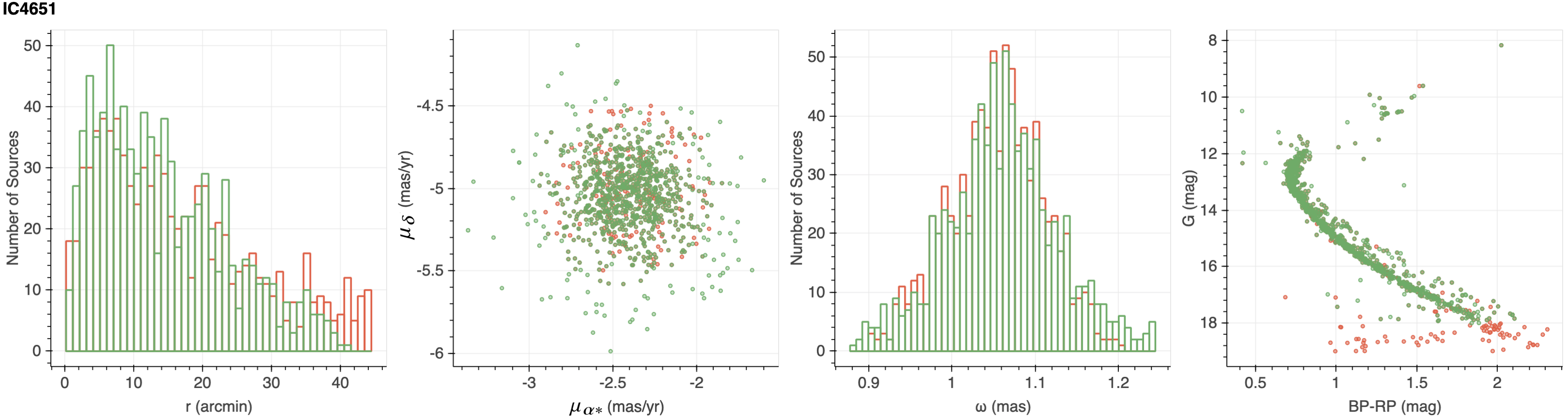}}
    \caption{Comparing the cluster member sources determined in this work with the members by CG18. The sources in red are extracted by ML-MOC. CG18 member sources (i.e. sources with membership probability greater than 0.5) are shown in green.}
\end{figure*}